\documentclass[journal]{IEEEtran}
\usepackage{cite}
\usepackage{graphicx}
\usepackage{textcomp}
\usepackage{xcolor}

\usepackage{soul}
\usepackage[mathscr]{eucal}
\usepackage[cmex10]{amsmath}
 \usepackage{amssymb,amsmath,amsthm,amsfonts,latexsym}
\usepackage{amsmath,graphicx,bm,xcolor,url,overpic}
\usepackage[caption=false]{subfig} 
\usepackage{fixltx2e}
\usepackage{array}
\usepackage{verbatim}
\usepackage{bm}

\usepackage{verbatim}
\usepackage{textcomp}
\usepackage{mathrsfs}
\usepackage{epstopdf}

\newcommand{\openone}{\leavevmode\hbox{\small1\normalsize\kern-.33em1}}

\catcode`~=11 \def\UrlSpecials{\do\~{\kern -.15em\lower .7ex\hbox{~}\kern .04em}} \catcode`~=13 

\allowdisplaybreaks[4]

\newcommand{\nn}{\nonumber}

\newcommand{\rma}{\mathrm{a}}

\newcommand{\rmb}{\mathrm{b}}

\newcommand{\rmd}{\mathrm{d}}

\newcommand{\rmE}{\mathrm{E}}

\newcommand{\rmp}{\mathrm{p}}

\newcommand{\rmt}{\mathrm{t}}

\newcommand{\rmu}{\mathrm{u}}

\newcommand{\rmw}{\mathrm{w}}

\DeclareMathAlphabet{\mathbsf}{OT1}{cmss}{bx}{n}
\DeclareMathAlphabet{\mathssf}{OT1}{cmss}{m}{sl}

\DeclareSymbolFont{bsfletters}{OT1}{cmss}{bx}{n}  
\DeclareSymbolFont{ssfletters}{OT1}{cmss}{m}{n}
\DeclareMathSymbol{\bsfGamma}{0}{bsfletters}{'000}
\DeclareMathSymbol{\ssfGamma}{0}{ssfletters}{'000}
\DeclareMathSymbol{\bsfDelta}{0}{bsfletters}{'001}
\DeclareMathSymbol{\ssfDelta}{0}{ssfletters}{'001}
\DeclareMathSymbol{\bsfTheta}{0}{bsfletters}{'002}
\DeclareMathSymbol{\ssfTheta}{0}{ssfletters}{'002}
\DeclareMathSymbol{\bsfLambda}{0}{bsfletters}{'003}
\DeclareMathSymbol{\ssfLambda}{0}{ssfletters}{'003}
\DeclareMathSymbol{\bsfXi}{0}{bsfletters}{'004}
\DeclareMathSymbol{\ssfXi}{0}{ssfletters}{'004}
\DeclareMathSymbol{\bsfPi}{0}{bsfletters}{'005}
\DeclareMathSymbol{\ssfPi}{0}{ssfletters}{'005}
\DeclareMathSymbol{\bsfSigma}{0}{bsfletters}{'006}
\DeclareMathSymbol{\ssfSigma}{0}{ssfletters}{'006}
\DeclareMathSymbol{\bsfUpsilon}{0}{bsfletters}{'007}
\DeclareMathSymbol{\ssfUpsilon}{0}{ssfletters}{'007}
\DeclareMathSymbol{\bsfPhi}{0}{bsfletters}{'010}
\DeclareMathSymbol{\ssfPhi}{0}{ssfletters}{'010}
\DeclareMathSymbol{\bsfPsi}{0}{bsfletters}{'011}
\DeclareMathSymbol{\ssfPsi}{0}{ssfletters}{'011}
\DeclareMathSymbol{\bsfOmega}{0}{bsfletters}{'012}
\DeclareMathSymbol{\ssfOmega}{0}{ssfletters}{'012}

\newtheorem{remark}{Remark}

\usepackage{booktabs}
\usepackage[ruled]{algorithm2e}
\begin{document}

\title{\scalebox{0.98}[1]{Exploiting UAV Attitude for Covert Communications:}\\Dynamics-Consistent Trajectory-Attitude Co-Design}
\author{Jinpeng~Xu, Lin Zhou, Xin Xie, Mengyuan Zhang and Jingjing Wang
\thanks{Jinpeng Xu is with the Department of Electronic and Computer Engineering, The Hong Kong University of Science and Technology, Hong Kong SAR (e-mail: eejinpengxu@ust.hk).}
\thanks{Lin Zhou is with the School of Automation and Intelligent Manufacturing,	Southern University of Science and Technology, Shenzhen 518000, China	(e-mail: zhoul9@sustech.edu.cn) (\emph{Corresponding author}).}
\thanks{Xin Xie is with the School of Communications and Information Engineering, Chongqing University of Posts and Telecommunications, Chongqing 400065, China (e-mail: xiex@cqupt.edu.cn). }
\thanks{Mengyuan Zhang and Jingjing Wang are with the School of Cyber Science and Technology, Beihang University, Beijing 100191, China (e-mail:\{myzhang18,drwangjj\}@buaa.edu.cn).}}
\maketitle

\begin{abstract}
Existing trajectory design for covert communications involving uncrewed aerial vehicles (UAVs) usually assume point-mass kinematics, which is overly idealized and overlooks the intrinsic coupling between flight maneuvers and antenna orientation. To address this issue, inspired by recent studies connecting UAV acceleration and UAV attitude, we study dynamics-consistent trajectory and attitude co-design for rotary-wing UAV covert communications. In our problem, a UAV equipped with a directional antenna transmits confidential information to a legitimate ground receiver, in the presence of a warden who attempts to detect UAV transmission. We explicitly model the acceleration-induced reduced attitude of the UAV and its impact on the attitude-dependent directional channel gain. To guarantee covertness, we derive a per-slot covert constraint using the Kullback--Leibler (KL) divergence metric for the warden. Subsequently, we formulate a trajectory and attitude co-design problem to maximize the average achievable covert rate, subject to the covertness constraint, UAV motion, thrust-magnitude feasibility, attitude-smoothness, and acceleration-domain safety constraints induced by roll and pitch angle limits of a UAV. The resulting problem is non-convex due to the coupled position-, acceleration-, and attitude-dependent channel gains. To solve the non-convex optimization problem, we propose a minorization--maximization (MM)-based algorithm, where a concave lower bound of the transmission rate and a convex upper bound of the covert constraint are calculated at each iteration. Furthermore, since practical UAV attitude control can be affected by actuator limitations, sensor noise, and control delays, we extend our proposed framework to imperfect attitude control and develop a robust design against attitude tracking errors. We have run extensive numerical examples, which confirm the effectiveness of our dynamics-consistent design and reveal the importance of jointly optimizing UAV trajectory and attitude for covert communications. In short, our study provides new perspectives and insights for characterizing the performance of covert communications with practical UAV systems.
\end{abstract}


\IEEEpeerreviewmaketitle
\vspace{-2mm}
\section{Introduction}
Uncrewed aerial vehicles (UAVs) are an important component of low-altitude wireless networks (LAWNs), owing to their high mobility, flexible deployment, and strong capability of establishing line-of-sight (LoS) air-to-ground links\cite{Xu2025FMFANET,Wang2025Consensus,xu2026sagsin_outage_capacity}. By acting as aerial base stations or mobile relays, UAVs can provide on-demand coverage and enhance communication services in dynamic and decentralized network environments where terrestrial infrastructure is unavailable, overloaded, or temporarily disrupted~\cite{xie2025sensing,xu2026service_efficiey,xie2025multiobjective}. However, the broadcast nature of wireless propagation and the high probability of LoS links make UAV communications vulnerable to surveillance by adversaries~\cite{Wang2025EDP,XJP_MIMOCovert}. This issue is particularly critical for security-sensitive applications, where the UAV transmission itself should remain hidden from wardens\cite{Hiding_inf_Bash_2015}. Therefore, UAV covert communication has attracted increasing attention as an effective technique to enable reliable information delivery while preventing the detection of wireless transmission by malicious third parties\cite{xu2024collaborative,He-CovertUAV-2026}.

\begin{figure}[!h]
    \centering\vspace{-1mm}
    \includegraphics[width=1\linewidth]
{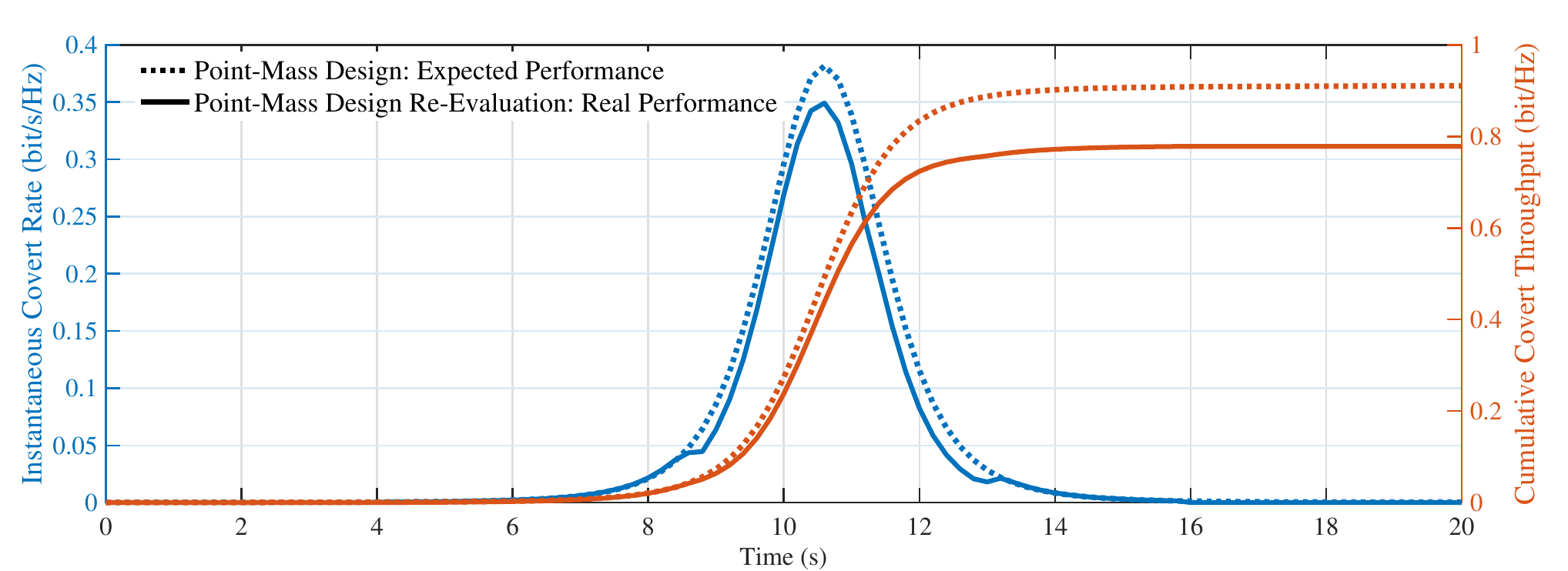}\vspace{-3mm}
\caption{Instantaneous covert rate and cumulative covert throughput in a three-node UAV covert communication scenario, where the UAV flies from its initial location to its final location while communicating with Bob in the presence of Willie. Dotted and solid curves denote the expected performance of the point-mass design and its re-evaluated real performance, respectively.}\centering\vspace{-0.5mm}
    \label{fig:point_mass_reevaluation}
\end{figure}

Given the high mobility of UAVs, trajectory optimization is critical and thus has been extensively studied for covert communication systems\cite{Zhou_TrajPower}. 
By properly designing the UAV trajectory, the transmitter UAV can approach the legitimate receiver, avoid the warden, or exploit favorable channel variations to improve the covert transmission rate. However, existing studies usually model the UAV as an ideal point-mass and focus on position and velocity design\cite{xu2024collaborative,Zhou_TrajPower}. This way, the antenna orientation is either ignored, assumed to be fixed, or treated as independently controllable. Such a simplification is convenient for optimization, but neglects a fundamental property of UAV communications: flight maneuvers, attitude, and antenna orientation are intrinsically coupled\cite{qiao_active_attitude_uav}.
To illustrate this issue, Fig.~\ref{fig:point_mass_reevaluation} compares the performance predicted by a conventional point-mass design with the real performance obtained by re-evaluating the same point-mass-designed trajectory using its corresponding UAV attitude parameters under the attitude-dependent communication model. 
The resulting performance degradation shows that neglecting UAV attitude can overestimate the achievable covert communication performance, highlighting the importance of explicitly accounting for attitude in covert communication design.

This performance mismatch stems from the intrinsic coupling between UAV motion, attitude, and antenna orientation. In practical rotary-wing UAVs, translational acceleration is generated by tilting the airframe\cite{Lee2010Geometric,Luo2024Synchronous}. Since onboard antennas are typically mounted on the UAV body, the antenna boresight varies with the UAV attitude during flight. As a result, trajectory maneuvers not only change the propagation distance and elevation angle, but also reshape the antenna gains toward both the legitimate receiver and the warden. This effect is particularly important for physical-layer security\cite{wiretap_1978} and becomes even more critical for covert communications\cite{Bashcovert2013}, where both the legitimate link and the warden's received power are highly sensitive to antenna orientation. A maneuver that improves the geometric distance to the legitimate user may unintentionally steer the communication beam main lobe toward the warden, failing to satisfy the covertness constraint. Conversely, properly exploiting the acceleration-induced attitude could help enhance the legitimate link while suppressing the warden's received power.
These observations indicate that UAV attitude should not be treated merely as a passive byproduct or an implementation detail. Instead, UAV attitude can serve as an additional design degree-of-freedom (DoF) for covert communications. 

Motivated by the above practical observation, we develop a dynamics-consistent trajectory and attitude co-design framework for rotary-wing UAV covert communications. In contrast to conventional point-mass trajectory designs, the framework explicitly captures the acceleration-attitude coupling and its impact on attitude-dependent directional transmission, which is of practical significance for the design and implementation of UAV covert communications. In the remainder, we briefly review existing trajectory-based studies on UAV covert communications, clarify our contributions beyond existing studies and state the organization of the rest of the paper.

 \vspace{-1mm}
\subsection{Related Studies}\vspace{-1mm}
The three-dimensional (3D)  mobility of UAVs offers an effective spatial DoF for covert communications, since trajectory, altitude, and hovering locations can be optimized to improve the legitimate link while reducing the warden's detection capability. Accordingly, trajectory-centric designs, often coupled with power allocation, beamforming, and cooperative jamming, have been widely studied in UAV covert communication systems.

Early efforts on UAV covert communications mainly considered joint deployment, trajectory, and transmit-power optimization under covertness constraints. In~\cite{A1,He-CovertUAV-2026}, UAV height and transmit power were jointly optimized to support covert transmission, which was later extended to geometric flying-location design in~\cite{A3,xu2024secret}. Along this direction, robust trajectory and power optimization were studied in~\cite{Zhou_TrajPower,jiang2021resource} by accounting for imperfect location of wardens. Besides horizontal trajectory design, vertical mobility has also been exploited. For example, altitude adaptation was used for energy-harvesting covert transmission in~\cite{A6}, while hovering-location and angular-domain designs were investigated for UAV-assisted internet-of-things (IoT) networks and energy-efficient covert transmission in~\cite{A8,CC9_angle}. 
Furthermore, more practical channel and transmission models have been considered in covert communication studies. The study in~\cite{CC7_nakagamiana} considered joint trajectory, power, and scheduling optimization for UAV covert communications under Nakagami fading channels, while finite-blocklength covert communication was investigated in UAV-assisted systems~\cite{B11} and, more recently, over quasi-static multiple-antenna fading channels~\cite{liu2026finite}.  Joint secrecy-and-covertness designs for UAV communications were further investigated in~\cite{xu2024collaborative,He-CovertUAV-2026}. Despite providing valuable insights, existing UAV-oriented studies generally model the UAV as a point mass, neglecting the intrinsic coupling between translational motion and attitude dynamics, thereby limiting their applicability to practical UAV systems.

As UAV covert communication moves from simple placement design to more dynamic network settings, trajectory optimization has been combined with different cooperative mechanisms. For artificial-noise-assisted transmission, the UAV trajectory and jamming strategy were jointly designed in~\cite{B7}. For relay-assisted covert communication, the trajectory design was coupled with blocklength allocation, relay forwarding, and user association in~\cite{B17,B18}. Related designs were also studied in IoT and vehicular communication scenarios, where trajectory, relay forwarding, jamming, and resource allocation were jointly optimized under uncertain warden locations~\cite{EE9_autonomousvehicular,B27_uavrelay}. In multi-user and multi-UAV systems, the trajectory was optimized together with bandwidth allocation, user association, and cooperative jamming to improve covert throughput and user fairness~\cite{B12,C3}. Furthermore, bidirectional air-ground covert communication was considered in~\cite{CC16_bidirectional}, where UAV trajectory, user association, and uplink/downlink power allocation were jointly optimized under bidirectional covert-detection constraints. More recently, clustered-device scenarios have been investigated, where a two-stage successive-hover-and-fly strategy was developed to reduce the complexity of joint inter-cluster trajectory and intra-cluster power optimization~\cite{CC5_devicecluster}. In addition to numerical optimization, several studies also derived structural properties of covert UAV trajectories, such as successive-hover-and-fly behavior~\cite{B12} and artificial-potential-field-based closed-form trajectories~\cite{B16}, which provide useful insights into mobility-aware covert transmission. On the other hand, learning-based UAV covert trajectory designs have also been studied, including deep reinforcement learning\cite{B10,wang2026local} and multi-agent reinforcement learning-based trajectory optimization methods~\cite{CC4_lightweightagent,C6,xu2026digital}. Nevertheless, these studies also generally adopt point-mass mobility models.

In summary, existing studies on UAV covert communications have substantially advanced mobility-enabled covert transmission by jointly designing UAV trajectories, deployment locations, transmit power, cooperative jamming, relaying strategies, user scheduling and association, and learning-based mobility control~\cite{A1,Zhou_TrajPower,A3,A8,CC9_angle,CC7_nakagamiana,B11,xu2024collaborative,He-CovertUAV-2026,B7,B17,B18,B27_uavrelay,EE9_autonomousvehicular,B12,C3,CC16_bidirectional,CC5_devicecluster,B16,B10,CC4_lightweightagent,xu2026digital,C6}. 
However, these studies generally rely on point-mass kinematics, where the attitude variations induced by UAV acceleration and maneuvering are neglected. 
As a result, they cannot characterize the coupling among UAV acceleration, attitude, antenna orientation, and air-ground channel gain, which could significantly affect both the legitimate transmission quality and the warden's detection capability.

On the other hand, attitude-aware UAV communication has been investigated in secrecy transmission~\cite{qiao_active_attitude_uav}. 
However, secrecy and covert communications pursue fundamentally different security objectives: the former aims to prevent the eavesdropper from decoding confidential information, whereas the latter aims to hide the very existence of wireless transmission. Therefore, the impact of dynamics-consistent UAV attitude on covert detection and covert communication performance remains unexplored, which motivates us to study trajectory-attitude co-design for UAV covert communications.

\subsection{Main Contributions}
To address the aforementioned research gaps, we propose a dynamics-consistent trajectory-attitude co-design framework for rotary-wing UAV covert communications. Different from conventional point-mass trajectory designs, our proposed framework explicitly captures the coupling between UAV flight maneuvers, acceleration-induced attitude, communication antenna gain, and covert communication performance. Our main contributions are summarized as follows.

\begin{enumerate}
\item To our best knowledge, this is the first study that embeds UAV attitude control into covert communication. Unlike existing trajectory-based UAV covert communication studies that model the UAV as an ideal point mass, we explicitly characterize the acceleration-induced attitude variation of a rotary-wing UAV and reveal its impact on both the legitimate channel and the warden's received power through attitude-dependent directional antenna gains.  As a result, our study reveals a new design DoF for covert communications.

\item To maximize the covert communication performance under practical flight constraints, we formulate a dynamics-consistent trajectory-attitude co-design problem that maximizes the average achievable rate at Bob subject to covertness, UAV kinematic, thrust-magnitude feasibility, attitude-smoothness, and acceleration-domain safety constraints. To solve the resulting non-convex problem, we develop a minorization-maximization (MM)-based algorithm that constructs a concave lower bound for the objective and a convex upper bound for the covert constraint at each iteration. 

\item To account for attitude-control imperfections in practical UAV systems caused by actuator limitations, sensor noise, and control delays, we further consider attitude tracking errors modeled by a bounded uncertainty set. Based on this model, we develop a robust trajectory-attitude design that accounts for worst-case attitude deviations and guarantees the covert communication requirement under imperfect attitude control.

\item Extensive numerical simulations are conducted to validate the effectiveness of the proposed framework. The results demonstrate that exploiting UAV attitude control can significantly improve covert communication performance over conventional point-mass and attitude-agnostic benchmark schemes, while confirming robustness under imperfect attitude control.

\end{enumerate}

\subsection{Organization of the Paper} 
The remainder of this paper is organized as follows. Section~\ref{sec:system_model} presents the system model, including the UAV dynamics-consistent attitude model and the attitude-dependent channel model. Section~\ref{sec:covert_analysis} analyzes the covert communication constraint based on KL divergence. Section~\ref{sec:problem_formulation} formulates the trajectory-attitude co-design problem and develops the proposed MM-based solution. Section~\ref{sec:imperfect_attitude_control} extends the framework to imperfect attitude control and presents the corresponding robust design. Numerical results are provided in Section~\ref{sec:simulation}, followed by conclusions in Section~\ref{sec:conclusion}.

\subsection*{Notations}
Vectors are denoted by boldface lower-case letters. Let $\mathbb{R}^{N}$ and $\mathbb{C}^{N}$ denote the spaces of $N$-dimensional real and complex vectors, respectively. We use ${\bf I}^N$ to denote an $N\times N$ identity matrix and $\bm{0}$ an all-zero vector or matrix of appropriate dimension. For a vector ${\bf x}$, $|{\bf x}|$ and ${\bf x}^T$ denote its Euclidean norm and transpose, respectively. The distribution of a circularly symmetric complex Gaussian random variable with mean $\mu$ and variance $\sigma^2$ is denoted by $\mathcal{CN}(\mu,\sigma^2)$, and $\sim$ means ``distributed as''. We use $\mathbb{E}[\cdot]$ to denote statistical expectation and $\mathcal{N}\triangleq{1,\ldots,N}$ the set of time-slot indices. For a differentiable function $f({\bf x})$, $\nabla_{\bf x} f({\bf x})$ denotes its gradient with respect to ${\bf x}$. Superscript $(r)$ denotes the value at the $r$-th iteration, while $\mathcal{O}(\cdot)$ denotes the standard big-O notation.

\section{System Model}
\label{sec:system_model}
\begin{figure}
\centering 
\includegraphics[width=1\linewidth]{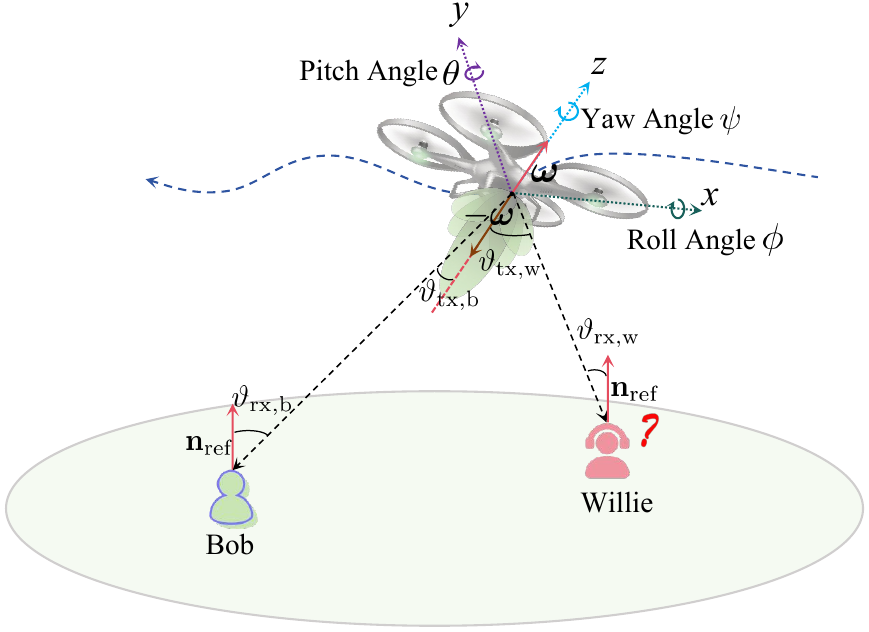}  
\caption{System model of dynamics-consistent UAV covert communications.} 
\label{fig:system_model}
\end{figure}

As shown in Fig. \ref{fig:system_model}, we consider a downlink covert communication system, where a rotary-wing UAV transmits confidential information to a legitimate ground receiver, Bob, in the presence of a warden, Willie, who attempts to detect the existence of the UAV transmission.
Bob is located at a fixed ground position, denoted by $\mathbf{p}_{\rmb}\in\mathbb{R}^3$, while the location of Willie is assumed to be known and denoted by $\mathbf{p}_{\rmw}\in\mathbb{R}^3$. The UAV operates in a 3D Cartesian coordinate system and is required to maintain a minimum altitude $H_{\min}$.
The UAV is dispatched from a prescribed initial location $\mathbf{p}_{\mathrm{I}}$ and is required to reach a designated final location $\mathbf{p}_{\mathrm{F}}$ within a mission duration $T$.
The mission duration $T$ is discretized into $N$ equally spaced time points
indexed by $\mathcal{N}\triangleq\{1,2,\ldots,N\}$, with time interval
$\delta_t=T/(N-1)$. Let $\mathbf{p}[n] = [x[n],y[n],z[n]]^T \in \mathbb{R}^3$ denote the UAV position at time slot $n$. The UAV velocity and acceleration are denoted by $\mathbf{v}[n] \in \mathbb{R}^3$ and $\mathbf{a}[n] \in \mathbb{R}^3$, respectively. Thus, the discrete-time kinematic evolution is
\begin{align}
\mathbf{p}[n+1]
&= \mathbf{p}[n] + \mathbf{v}[n]\delta_t
+ \tfrac{1}{2}\mathbf{a}[n]\delta_t^2,~\forall n \in \mathcal{N}\setminus\{N\}, \label{eq:kinematics_p} \\[1mm]
\mathbf{v}[n+1]
&= \mathbf{v}[n] + \mathbf{a}[n]\delta_t,~\forall n \in \mathcal{N}\setminus\{N\}. \label{eq:kinematics_v}
\end{align}

The UAV motion is subject to the following constraints
\begin{align}
\|\mathbf{v}[n]\| 
\le& V_{\max}, ~  \forall n \in \mathcal{N},\\[0.5mm]
\|\mathbf{a}[n]\| \le& A_{\max},~  \forall n \in \mathcal{N}, \\[0.5mm]
z[n] ~\ge& H_{\min},~  \forall n \in \mathcal{N}.
\end{align}
\subsection{Dynamics-Consistent Attitude Model}
In a rotary-wing UAV, the attitude of the airframe is intrinsically coupled with its translational acceleration due to rigid-body dynamics\cite{qiao_active_attitude_uav}.  
Let $\boldsymbol{\omega}[n] \in \mathbb{R}^3$ denote the unit vector along the UAV body vertical axis. Since the onboard directional antenna is mounted downward, its boresight direction is given by $-\boldsymbol{\omega}[n] \in \mathbb{R}^3$. 
Physically, $\boldsymbol{\omega}[n]$ represents the instantaneous body vertical axis of the UAV in the inertial frame and determines the pointing direction of the directional antenna. Specifically, it can be expressed in terms of the roll, pitch, and yaw angles $(\phi[n], \theta[n], \psi[n])$ through the standard rotation matrix as\cite{qiao_active_attitude_uav}
\begin{equation}
\boldsymbol{\omega}[n] =
\begin{bmatrix}
\cos\psi[n]\sin\theta[n]\cos\phi[n] + \sin\psi[n]\sin\phi[n] \\
\sin\psi[n]\sin\theta[n]\cos\phi[n] - \cos\psi[n]\sin\phi[n] \\
\cos\theta[n]\cos\phi[n]
\end{bmatrix},
\label{eq:UAV_body_z_axis}
\end{equation}
where $\phi[n]$, $\theta[n]$, and $\psi[n]$ denote the roll, pitch, and yaw angles, respectively.
According to Newton--Euler rigid-body dynamics and previous study \cite{qiao_active_attitude_uav}, the UAV body vertical axis, i.e., the reduced attitude, is uniquely determined by the required translational acceleration:
\begin{align}
\boldsymbol{\omega}[n] &= \frac{\mathbf{a}[n] + \mathbf{g}}{\|\mathbf{a}[n] + \mathbf{g}\|},
\label{eq:attitude_coupling}
\end{align}
where $\mathbf{a}[n] \in \mathbb{R}^3$ is the UAV acceleration at time slot $n$, and $\mathbf{g} = [0,0,g]^T$ denotes gravitational acceleration. 
This indicates that the antenna orientation cannot be directly controlled, but is adjusted indirectly via UAV maneuvers.

To ensure physically realizable and safe flight, the UAV acceleration vector is subject to convex safety constraints that capture the allowable angular limits of the airframe.  
Let $\mathbf{a}_g[n] = \mathbf{a}[n] + \mathbf{g} = [a_x[n],a_y[n],a_z[n]+ g]^T$ denote the net acceleration vector. 
Since yaw primarily determines the UAV heading and is not involved in the acceleration-induced reduced-attitude mapping, we do not optimize $\psi[n]$ explicitly. Instead, $\psi[n]$ is treated as a prescribed heading profile provided by a high-level guidance module.
For a given $\psi[n]$, the longitudinal and lateral acceleration components with respect to the heading direction are defined as
\begin{equation}
\begin{bmatrix}
a_\parallel[n] \\
a_\perp[n]
\end{bmatrix}
:=
\begin{bmatrix}
\cos\psi[n] & \sin\psi[n] \\
-\sin\psi[n] & \cos\psi[n]
\end{bmatrix}
\begin{bmatrix}
a_x[n] \\
a_y[n]
\end{bmatrix}.
\label{eq:apar_aperp}
\end{equation}
For all $n\in\mathcal N$, the safety inequalities are then formulated as the following conservative acceleration-domain constraints:
\begin{subequations}
\begin{align}
(a_z[n]+g)\tan\phi_{\max} 
&\ge|a_\perp[n]|
, \\
a_\parallel[n] - (a_z[n]+g)\tan\theta_{\min}
&\ge 0, \\
a_\parallel[n] - (a_z[n]+g)\tan\theta_{\max}
&\le 0, \\
a_z[n] + g
&\ge 0.
\end{align}
\end{subequations}

\subsection{Dynamics-Consistent Rician Channel Model}
\label{sec:channel_model}

In our covert communication scenario, the UAV serves a legitimate ground receiver Bob while avoiding detection by the warden Willie. 
For node $\dagger\in\{\rmb,\rmw\}$, we define $\mathbf d_\dagger[n] \triangleq \mathbf p[n]-\mathbf p_\dagger$ and $d_\dagger[n]\triangleq \|\mathbf d_\dagger[n]\|$.
The vertical reference axis of the ground antenna is denoted by $\mathbf n_{\rm ref}=[0,0,1]^T$.
Following the off-axis angle convention, the off-boresight angle $\vartheta_{{\rm tx},\dagger}[n]$ between the UAV antenna boresight and the LoS direction toward node $\dagger$ and the off-axis angle $\vartheta_{{\rm rx},\dagger}[n]$ at the ground receiver are respectively defined as
\begin{align}
\vartheta_{{\rm tx},\dagger}[n]
&=
\arccos\!\left(
\frac{\mathbf d_\dagger^T[n]\boldsymbol{\omega}[n]}{d_\dagger[n]}
\right),
\label{eq:theta_tx_rician}
\\
\vartheta_{{\rm rx},\dagger}[n]
&=
\arccos\!\left(
\frac{\mathbf d_\dagger^T[n]\mathbf n_{\rm ref}}{d_\dagger[n]}
\right).
\label{eq:theta_rx_rician}
\end{align}
Since $\mathbf d_\dagger[n]=\mathbf p[n]-\mathbf p_\dagger$ points from node $\dagger$ to the UAV, the above definitions are consistent with the downward-facing UAV antenna
boresight $-\boldsymbol{\omega}[n]$ and the upward-facing ground reference axis $\mathbf n_{\rm ref}$.

Consistent with\cite{XJP_Model,xu2026sagsin_outage_capacity}, we model the UAV-to-ground channel as a Rician fading channel consisting of a deterministic LoS component and an aggregated non-line-of-sight (NLoS) component. Let $\kappa_\dagger\ge 0$ denote the Rician factor associated with node $\dagger$. Then, the corresponding Rician-averaged geometric channel gain is given by 
\begin{align}
\hat F_\dagger[n]
=
\frac{\kappa_\dagger}{1+\kappa_\dagger}
\bar F_\dagger[n]
+
\frac{1}{1+\kappa_\dagger}
\tilde F_\dagger[n],
\label{eq:Fhat_Rician}
\end{align}
where the LoS component $\bar F_{\dagger}[n]$ is explicitly controlled by the UAV attitude through the antenna radiation pattern, whereas the NLoS component $\tilde F_{\dagger}[n]$ represents the averaged residual scattered energy. Both the LoS component $\bar F_\dagger[n]$ and the NLoS component $\tilde F_\dagger[n]$ are specified as follows. 
\subsubsection{Attitude-Dependent LoS Gain} Based on 3GPP TR 38.901 \cite[Table 7.3-1]{3gpp38901}, we adopt a peak-normalized single-element directional antenna pattern to model
the attitude-dependent LoS gain. Specifically, the antenna gain as a function
of the off-boresight angle $\vartheta$ is given by 
\begin{equation}
G(\vartheta)
=
10^{-\frac{1}{10}
\min\left\{
12\left(\frac{\vartheta}{\vartheta_{3{\rm dB}}}\right)^2,
{\rm SLA}
\right\}},
\label{eq:antenna_gain_pattern} 
\end{equation}
where $\vartheta_{3{\rm dB}}$ denotes the half-power beamwidth and ${\rm SLA}$
denotes the maximum side-lobe attenuation. By construction, $G(0)=1$.
Accordingly, the attitude-dependent LoS geometric gain from the UAV to node
$\dagger$ is defined as
\begin{equation}
\bar F_\dagger[n]
=
\frac{
G\!\left(\vartheta_{{\rm tx},\dagger}[n]\right)
G\!\left(\vartheta_{{\rm rx},\dagger}[n]\right)
}
{d_\dagger^2[n]}.
\label{eq:Fbar_LoS}
\end{equation}
The term $\bar F_\dagger[n]$ captures the LoS free-space distance attenuation and the
directional antenna responses at both the UAV transmitter and the ground receiver. In particular, the UAV attitude affects the LoS channel through $\vartheta_{{\rm tx},\dagger}[n]$, while $\vartheta_{{\rm rx},\dagger}[n]$ is determined by the ground receiver orientation and the UAV-node geometry.

\subsubsection{Averaged NLoS Gain}

Besides the deterministic LoS component, the UAV-to-ground channel may contain reflected, scattered, and diffracted components. Since the instantaneous NLoS
scatterers are environment-dependent, we model their aggregate effect statistically.
For node $\dagger$, suppose that the NLoS propagation consists of $M$ scattering clusters, indexed by $m\in\{1,\ldots,M\}$, and each cluster contains $I$ unresolved sub-paths, indexed by $i\in\{1,\ldots,I\}$.
For the $i$-th sub-path in the $m$-th cluster, let
$\tilde\vartheta_{{\rm tx},\dagger,m,i}[n]$ and
$\tilde\vartheta_{{\rm rx},\dagger,m,i}[n]$ denote the transmit-side and receive-side off-axis angles, respectively. These angles are determined by the corresponding scatterer geometry rather than by the direct LoS vector.

The normalized angular-scattering power contribution of the $i$-th NLoS
sub-path in the $m$-th cluster is modeled as
\begin{equation}
\!\!q_{\dagger,m,i}[n]
\!=\!
\beta_{\dagger,m,i}[n]\,
G\!\left(\tilde\vartheta_{{\rm tx},\dagger,m,i}[n]\right)
G\!\left(\tilde\vartheta_{{\rm rx},\dagger,m,i}[n]\right),
\label{eq:NLoS_subpath_power}
\end{equation}
where $\beta_{\dagger,m,i}[n]\ge 0$ is the normalized reflection/scattering power coefficient of the corresponding sub-path, and $G(\cdot)$ is the peak-normalized antenna pattern defined in \eqref{eq:antenna_gain_pattern}. Furthermore, the aggregate NLoS small-scale coefficient can be written as\cite{XJP_Model}
\begin{equation}
\tilde h_{\dagger}[n]
=
d_\dagger^{-\frac{\tilde\alpha}{2}}[n]
\sum_{m=1}^{M}
\sum_{i=1}^{I}
\sqrt{q_{\dagger,m,i}[n]}\,
e^{j\Phi_{\dagger,m,i}[n]},
\label{eq:NLoS_sum_model}
\end{equation}
where $\tilde\alpha>2$ is the NLoS path-loss exponent, and
$\Phi_{\dagger,m,i}[n]$ is the random phase of the corresponding sub-path.
We assume that the NLoS sub-paths have independent phases or are resolvable over delay, such that the cross terms vanish after ensemble averaging. Therefore, the averaged NLoS power is
\begin{align}
\!\!\!\!\mathbb E_{\mathcal S_\dagger[n]}\!
\left\{\!
\left|\tilde h_{\dagger}[n]\right|^2\!
\right\}
\!=&
d_\dagger^{-\tilde\alpha}[n]\,
\mathbb E_{\mathcal S_\dagger[n]}
\left[
\sum_{m=1}^{M}
\sum_{i=1}^{I}
q_{\dagger,m,i}[n]
\right]
\\
=&
d_\dagger^{-\tilde\alpha}[n]\,
\mathbb E_{\mathcal S_\dagger[n]}
\Bigg[
\sum_{m=1}^{M}
\sum_{i=1}^{I}
\beta_{\dagger,m,i}[n]\nn\\
&\!\times\! G\!\left(\tilde\vartheta_{{\rm tx},\dagger,m,i}[n]\right)\!
G\!\left(\tilde\vartheta_{{\rm rx},\dagger,m,i}[n]\right)\!\!
\Bigg]\!,\!\!
\label{eq:NLoS_average_power}
\end{align}
where the random scattering state is defined as
{\small\begin{equation}
\!\!\mathcal S_\dagger[n]
\!\!\triangleq\!\!
\left\{\!
\beta_{\dagger,m,i}[n],\!
\tilde\vartheta_{{\rm tx},\dagger,m,i}[n],\!
\tilde\vartheta_{{\rm rx},\dagger,m,i}[n],\!
\Phi_{\dagger,m,i}[n]\!
\right\}_{\!m=1,i=1}^{\!M,I}\!.\!\!
\label{eq:scattering_state_simple}
\end{equation}}
Accordingly, the effective NLoS scattering coefficient is
{\small\begin{align}
\!\!\!\!\tilde{\eta}_{\dagger}[n] \!\!
\triangleq\! 
\mathbb E_{\mathcal S_\dagger[n]}\!\!\!
\left[ 
\sum_{m=1}^{M}\!
\sum_{i=1}^{I}\!
\beta_{\dagger,m,i}[n] 
G\!\left(\!\tilde\vartheta_{{\rm tx},\dagger,m,i}[n]\!\right)\!
G\!\left(\!\tilde\vartheta_{{\rm rx},\dagger,m,i}[n]\!\right)
\!\!\right]\!.\!\!
\label{eq:eta_N_definition}
\end{align}}
Thus, the averaged NLoS geometric gain is
\begin{equation}
\tilde F_\dagger[n]
\triangleq
\mathbb E_{\mathcal S_\dagger[n]}
\left\{
\left|\tilde h_{\dagger}[n]\right|^2
\right\}
=
\tilde{\eta}_{\dagger}[n]  d_\dagger^{-\tilde\alpha }[n].
\label{eq:Ftilde_NLoS}
\end{equation}
Therefore, $\tilde F_\dagger[n]$ is the variance, or equivalently the averaged power, of the aggregate NLoS component.

\subsubsection{Received Signals at Bob and Willie}
Without loss of generality, we assume identical noise power at Bob and Willie, i.e.,
$\sigma_{\rmb}^2=\sigma_{\rmw}^2=\sigma^2$, and define the reference SNR as  
\begin{equation}
\rho_0
\triangleq
\frac{P_{\rmt}G_0^2}{\sigma^2}
\left(\frac{\lambda}{4\pi}\right)^2,
\label{eq:rho0_rician}  
\end{equation}
where $P_{\rmt}$ is the UAV transmit power, $G_0$ is the peak antenna gain, and $\lambda$ is the carrier wavelength. 
Accordingly, the averaged received SNR at node $\dagger$ is  
\begin{equation}
\gamma_\dagger[n]
=
\rho_0 \hat F_{\dagger}[n].
\label{eq:gamma_k_rician}  
\end{equation}

Let $x[n]$ denote the normalized transmitted symbol with $\mathbb E\{|x[n]|^2\}=1$. Then, the received signal at Bob is given by
\begin{equation}
y_{\rmb}[n]
=
\sqrt{
P_{\rmt}G_0^2
\hat F_{\rmb}[n]
\left(\frac{\lambda}{4\pi}\right)^2
}
x[n]
+
n_{\rmb}[n],
\label{eq:yb_rician} 
\end{equation}
where $n_{\rmb}[n]\sim\mathcal{CN}(0,\sigma^2)$ is AWGN. 
Therefore, the achievable rate at Bob is given by  
\begin{equation}
R_{\rmb}[n]
=
\log_2
\left(
1+
\rho_0\hat F_{\rmb}[n]
\right).
\label{eq:Rb_rician}  
\end{equation}

From Willie's perspective, at each time slot, Willie performs a binary hypothesis test to decide whether
the UAV is transmitting. Specifically, $\mathcal H_0$ corresponds to the case where the UAV remains silent, while $\mathcal H_1$ corresponds to the case where the UAV transmits. Under the two hypotheses, the received signal at Willie is given by
\begin{equation}
y_{\rmw}[n]
=
\begin{cases}
n_{\rmw}[n], & \mathcal H_0,\\[1mm]
\sqrt{
P_{\rmt}G_0^2
\hat F_{\rmw}[n]
\left(\frac{\lambda}{4\pi}\right)^2
}
\,x[n]
+
n_{\rmw}[n],
& \mathcal H_1,
\end{cases}
\label{eq:willie_signal_rician}
\end{equation}
where $n_{\rmw}[n]\sim\mathcal{CN}(0,\sigma_{\rmw}^2)$ denotes AWGN at Willie.
Accordingly, under hypothesis $\mathcal H_1$, the instantaneous received SNR at Willie is given by  
\begin{equation}
\gamma_{\rmw}[n]
=
\frac{
P_{\rmt}G_0^2
\hat F_{\rmw}[n]
\left(\frac{\lambda}{4\pi}\right)^2
}
{\sigma_{\rmw}^2}.
\label{eq:gamma_w_general} 
\end{equation}
Based on the received signal models under $\mathcal H_0$ and $\mathcal H_1$, we next characterize Willie's detection performance and derive the corresponding covert communication constraint.

\section{Analysis of the Covertness Measure}
\label{sec:covert_analysis}

In this section, we adopt the Kullback--Leibler (KL) divergence as the covertness measure and derive the corresponding characterization in the presence of the warden Willie.

We assume Gaussian signaling with $x[n,\ell]\sim\mathcal{CN}(0,1)$ independently across symbols.
At each time slot $n \in \mathcal{N}$, Willie collects a length-$L$ observation sequence to determine whether the UAV is transmitting. 
Let $\mathbf{y}_\rmw[n] \triangleq \{y_\rmw[n,\ell]\}_{\ell=1}^{L}$ denote the received signal sequence at Willie in time slot $n$, where $\ell$ is the symbol index within slot $n$. 
Accordingly, Willie performs a binary hypothesis test in each time slot based on $\mathbf{y}_\rmw[n]$.
Let $\mathbb{P}_{0,n}$ and $\mathbb{P}_{1,n}$ denote the joint distributions of $\mathbf{y}_\rmw[n]$ under hypotheses $\mathcal{H}_0$ and $\mathcal{H}_1$, respectively. 
Assuming that the $L$ observations within each time slot are independent and identically distributed, we have  
\begin{equation}
\begin{cases}
\displaystyle \mathbb{P}_{0,n} = \prod_{\ell=1}^{L} f(y_\rmw[n,\ell]|\mathcal{H}_0),\\[2mm]
\displaystyle \mathbb{P}_{1,n} = \prod_{\ell=1}^{L} f(y_\rmw[n,\ell]|\mathcal{H}_1),
\end{cases}  
\end{equation}
where $f(y_\rmw[n,\ell]|\mathcal{H}_j)$ is the probability density function (PDF) under hypothesis $\mathcal{H}_j$, $j \in \{0,1\}$.
Since $y_\rmw[n,\ell]$ follows zero-mean complex Gaussian distributions under both hypotheses, we further have
\begin{equation}
f(y_\rmw[n,\ell]) =
\begin{cases}
\displaystyle \frac{1}{\pi \sigma_0^2[n]} \exp\!\left(-\frac{|y_\rmw[n,\ell]|^2}{\sigma_0^2[n]}\right), & \mathcal{H}_0,\\[2.5mm]
\displaystyle \frac{1}{\pi \sigma_1^2[n]} \exp\!\left(-\frac{|y_\rmw[n,\ell]|^2}{\sigma_1^2[n]}\right), & \mathcal{H}_1,
\end{cases}
\end{equation}
where $\sigma_0^2[n] = \sigma^2$ and $\sigma_1^2[n] = \sigma^2\big(1+\gamma_\rmw[n]\big)$.

We adopt the KL divergence as the covertness metric to characterize the distinguishability between $\mathcal H_0$ and $\mathcal H_1$ at Willie. By exploiting the additivity of KL-divergence, the KL-divergence in time slot $n$ is\cite{XJP_MIMOCovert}  
\begin{align}
\!\!\!\mathcal{D}[n]
&\triangleq \mathcal{D}(\mathbb{P}_{0,n}\|\mathbb{P}_{1,n}) \\*
&= \!\sum_{\ell=1}^{L} \!\int f(y_\rmw[n,\ell]|\mathcal{H}_0)
\ln \frac{f(y_\rmw[n,\ell]|\mathcal{H}_0)}{f(y_\rmw[n,\ell]|\mathcal{H}_1)}
\rmd y_\rmw[n,\ell] \\*
&= L\left[\ln\big(1+\gamma_\rmw[n]\big) - \frac{\gamma_\rmw[n]}{1+\gamma_\rmw[n]}\right].
\end{align}

In typical covert communication scenarios, the UAV transmit power is low and the received SNR at Willie satisfies $\gamma_\rmw[n] \ll 1$. 
Using the Taylor expansions
\begin{align}
\ln(1+\gamma_\rmw[n]) &= \gamma_\rmw[n] - \frac{\gamma_\rmw^2[n]}{2} + O(\gamma_\rmw^3[n]),\\
\frac{\gamma_\rmw[n]}{1+\gamma_\rmw[n]} &= \gamma_\rmw[n] - \gamma_\rmw^2[n] + O(\gamma_\rmw^3[n]),
\end{align}
we obtain
\begin{align}
\mathcal{D}[n]
&= L\left[\ln\big(1+\gamma_\rmw[n]\big) - \frac{\gamma_\rmw[n]}{1+\gamma_\rmw[n]}\right]
\\& \approx \frac{L}{2}\gamma_\rmw^2[n].
\end{align}
Therefore, in the low-SNR regime, the per-slot KL-divergence can be conservatively upper-bounded by
\begin{equation}
\mathcal{D}[n]
\lesssim
\frac{L}{2}\big(\rho_0 \hat F_{\rmw}(\mathbf{d}_\rmw[n], \boldsymbol{\omega}[n])\big)^2.
\end{equation}
To ensure per-slot covertness, we impose
\begin{equation}
\mathcal{D}[n] \le \epsilon, \qquad \forall n \in \mathcal{N}.
\end{equation}
Accordingly, under the low-SNR approximation, the covert communication constraint can be expressed as
\begin{equation} \label{covert_measure}
\rho_0 \hat F_{\rmw}(\mathbf{d}_\rmw[n], \boldsymbol{\omega}[n]) \le \sqrt{\frac{2\epsilon}{L}}, \qquad \forall n \in \mathcal{N}.
\end{equation}
\begin{remark}
The approximation $\mathcal{D}[n] \approx \frac{L}{2}\gamma_\rmw^2[n]$ is asymptotically exact when $\gamma_\rmw[n]\to 0$. In practice, for finite but small $\gamma_\rmw[n]$, the quadratic approximation provides a conservative upper bound on the exact KL-divergence. Thus, the covert constraint is slightly stricter than the original one, which improves covertness at the cost of some performance loss.
\end{remark}

\section{Problem Formulation and Solutions}
\label{sec:problem_formulation}
In this section, we formulate the covert communication performance maximization problem under UAV maneuver and attitude constraints. To address the resulting non-convexity, an iterative MM framework is subsequently developed to obtain a locally optimal solution.

\subsection{Problem Formulation}
Our objective is to optimize the UAV trajectory and attitude to maximize the covert transmission rate to Bob while satisfying the covertness constraints at Willie and all physical flight constraints. 
 Let $\mathbf{P} := \{\mathbf{p}[n]\}_{n=1}^{N}\in\mathbb{R}^{3\times N}$, $\mathbf{V} := \{\mathbf{v}[n]\}_{n=1}^{N}\in\mathbb{R}^{3\times N}$, $\mathbf{A} := \{\mathbf{a}[n]\}_{n=1}^{N}\in\mathbb{R}^{3\times N}$, and $\boldsymbol{\Omega} := \{\boldsymbol{\omega}[n]\}_{n=1}^{N}\in\mathbb{R}^{3\times N}$. Thus, the corresponding optimization problem can be formulated as
\begin{subequations}
\begin{align}
\!\!\!\!\text{(P1)}
\max_{\mathbf{P},\mathbf{V},\mathbf{A},\boldsymbol{\Omega}}
&
\frac{1}{N}\sum_{n=1}^{N}
\log_2\!\Big(1+\rho_0 \hat F_{\rmb}(\mathbf{d}_\rmb[n],\boldsymbol{\omega}[n])\Big)\label{p0_objfun}
\\
\text{s.t.}\quad
&
\mathbf{p}[1]=\mathbf{p}_{\mathrm{I}},\quad
\mathbf{p}[N]=\mathbf{p}_{\mathrm{F}},\label{p0_start_final}
\\
&
\mathbf{v}[1]=\mathbf{0},\quad
\mathbf{v}[N]=\mathbf{0},\label{p0_v0_vN}
\\
&{\small\!
\mathbf{p}[n\!\!+\!\!1\!]\!\!=\!\!\mathbf{p}[n]\!\!+\!\mathbf{v}[n]\delta_t\!\!+\!\!\frac{1}{2}\mathbf{a}[n]\delta_t^2, \!\forall n \!\in\!\mathcal{N}\!\setminus\!\!\{\!N\!\}},\!\!\!\label{p0_cons_p}
\\
&
\mathbf{v}[n\!+\!1]\!=\!\mathbf{v}[n]\!+\!\mathbf{a}[n]\delta_t,
\forall n \in \mathcal{N}\!\setminus\!\{\!N\!\},\label{p0_cons_v_a}
\\
&
\|\mathbf{v}\![n]\| \le  V_{\max},
\|\mathbf{a}[n]\| \le A_{\max},
\forall n \in \mathcal{N},\label{p0_cons_vaz}
\\
& H_{\max}\ge z[n] \ge H_{\min},\forall n \in \mathcal{N},\label{p0_cons_high}
\\
&
\rho_0 \hat F_{\rmw}(\mathbf{d}_\rmw[n],\boldsymbol{\omega}[n])
\le
\sqrt{\frac{2\epsilon}{L}},
\forall n\in\mathcal{N},\label{p0_covertness}
\\
&
\boldsymbol{\omega}[n]
=
\dfrac{\mathbf{a}[n]+\mathbf{g}}
{\|\mathbf{a}[n]+\mathbf{g}\|},
\quad \forall n\in\mathcal{N},\label{p0_cons_w}
\\
&
\|\mathbf{a}[n]+\mathbf{g}\|\le \dfrac{T_{\max}}{m_\rmu},
\quad \forall n\in\mathcal{N},\label{p0_cons_sec}
\\
&
\|\boldsymbol{\omega}[n]\|=1,
\quad \forall n\in\mathcal{N},\label{p0_consw1}
\\
&
\|\boldsymbol{\omega}[n\!+\!1]\!-\!\boldsymbol{\omega}[n]\|^2\!\le \!\delta_{\rm rot},
\forall n \!\in\! \mathcal{N}\!\setminus\!\{\!N\!\},\label{p0_cons_secw}
\\
&
(a_z[n]+g)\tan\phi_{\max}
\ge|a_\perp[n]|
, \forall n\in\mathcal N,\label{p0_cons_a1}
\\
&
a_\parallel[n]\!-\!(a_z[n]\!\!+\!g)\tan\theta_{\min}\!\ge \!0, \forall n\!\in\!\mathcal{N},\label{p0_cons_a2}
\\
&
a_\parallel[n]\!-\!(a_z[n]\!\!+\!g)\!\tan\theta_{\max}\!\le\! 0, \forall n\!\in\!\mathcal{N},\label{p0_cons_a3}
\\
&
a_z[n]+g\ge0,
\quad \forall n\in\mathcal{N}.\label{p0_cons_a4}
\end{align}
\end{subequations}
In problem (P1), constraints \eqref{p0_start_final} and \eqref{p0_v0_vN} specify the initial/final positions of the UAV. Constraints \eqref{p0_cons_p} and \eqref{p0_cons_v_a} characterize the discrete-time UAV kinematics, while \eqref{p0_cons_vaz} imposes the maximum velocity, maximum acceleration, and minimum-altitude requirements. The covert communication requirement at Willie is enforced by \eqref{p0_covertness}. Constraint \eqref{p0_cons_w} captures the dynamics-consistent acceleration-attitude coupling, and \eqref{p0_cons_sec} represents the maximum thrust constraint. Constraints \eqref{p0_consw1} and \eqref{p0_cons_secw} impose the unit-norm and inter-slot smoothness requirements on the UAV attitude, respectively. Finally, \eqref{p0_cons_a1}--\eqref{p0_cons_a4} describe the acceleration-domain safety constraints induced by the roll and pitch angle limits.  

Note that problem (P1) is highly non-convex. In particular, the Rician-averaged channel gains depend nonlinearly on the UAV position and attitude, while the attitude is further coupled with the acceleration through the normalized mapping in \eqref{p0_cons_w}. Consequently, both the objective and covertness constraint involve coupled non-convex functions of the trajectory and acceleration. Furthermore, the unit-norm constraint in \eqref{p0_consw1} and the attitude-smoothness constraint in \eqref{p0_cons_secw} result in a non-convex feasible set. These couplings prevent direct application of conventional trajectory optimization or alternating optimization methods, since fixing one variable block need not yield a convex subproblem. Therefore, problem (P1) cannot be directly solved by standard convex optimization tools\cite{cvx}, which motivates the MM-based reformulation developed next.

\subsection{Proposed Solution}
\label{sec:solution}
To address the non-convexity of (P1), we develop an MM-based algorithm. Specifically,  our algorithm proceeds in four steps: 1) the attitude variable is eliminated using the dynamics-consistent acceleration-attitude mapping, leading to reduced position- and acceleration-dependent channel gains; 2) the average covert transmission rate objective is replaced by its local concave minorizer; 3) the covert and angular-smoothness constraints are conservatively approximated by convex upper bounds; 4) the resulting convex subproblem is solved within a trust region and updated until convergence. The validity of the  algorithm is ensured by the MM-based lower- and upper-bound constructions within the trust region, which preserve feasibility and guarantee monotonic objective improvement.

\subsubsection{Attitude Elimination and Reduced Channel Reformulation}
\label{subsec:attitude_elimination}

Based on \eqref{eq:attitude_coupling}, we eliminate the attitude variable $\boldsymbol{\omega}$ by using the dynamics-consistent mapping.
For node $\dagger$, the corresponding off-axis angles after attitude
elimination are
\begin{align}
\vartheta_{{\rm tx},\dagger}(\mathbf p[n],\mathbf a[n])
&=
\arccos\!\left(
\frac{\mathbf d_\dagger^T[n]\boldsymbol{\omega}[n]}{d_\dagger[n]}
\right),
\label{eq:theta_tx_reduced}
\\
\vartheta_{{\rm rx},\dagger}(\mathbf p[n])
&=
\arccos\!\left(
\frac{\mathbf d_\dagger^T[n]\mathbf n_{\rm ref}}{d_\dagger[n]}
\right).
\label{eq:theta_rx_reduced}
\end{align}
Accordingly, the reduced LoS gain is given by
\begin{equation}
\bar F_\dagger(\mathbf p[n],\mathbf a[n])
=
\frac{
G\!\left(\vartheta_{{\rm tx},\dagger}(\mathbf p[n],\mathbf a[n])\right)
G\!\left(\vartheta_{{\rm rx},\dagger}(\mathbf p[n])\right)
}
{d_\dagger^2[n]}.
\label{eq:Fbar_reduced}
\end{equation}
Therefore, the
Rician-averaged reduced channel gain is
\begin{equation}
\!\!\hat F_\dagger(\mathbf p[n],\mathbf a[n])
\!=\!
\frac{\kappa_\dagger}{1\!+\!\kappa_\dagger}
\bar F_\dagger(\mathbf p[n],\mathbf a[n])
\!+\!
\frac{1}{1\!+\!\kappa_\dagger}
\tilde F_\dagger(\mathbf p[n]).
\label{eq:Fhat_reduced}
\end{equation}

The gradients required by the MM procedure follow from the chain rule. Define
$\mathbf s[n]=\mathbf a[n]+\mathbf g$. The Jacobian of
$\boldsymbol{\omega}[n]=\mathbf s[n]/\|\mathbf s[n]\|$ is given by
\begin{equation}
\mathbf J_{\omega}[n]
=
\frac{1}{\|\mathbf s[n]\|}
\left(
\mathbf I_3
-
\frac{\mathbf s[n]\mathbf s^T[n]}{\|\mathbf s[n]\|^2}
\right).
\label{eq:Jomega_mm}
\end{equation}
Then, we further have
\begin{align}
\nabla_{\mathbf p}\hat F_\dagger
&=
\frac{\kappa_\dagger}{1+\kappa_\dagger}
\nabla_{\mathbf p}\bar F_\dagger
+
\frac{1}{1+\kappa_\dagger}
\nabla_{\mathbf p}\tilde F_\dagger,
\label{eq:grad_p_Fhat_reduced}
\\
\nabla_{\mathbf a}\hat F_\dagger
&=
\frac{\kappa_\dagger}{1+\kappa_\dagger}
\mathbf J_{\omega}^T[n]
\nabla_{\boldsymbol{\omega}}\bar F_\dagger.
\label{eq:grad_a_Fhat_reduced}
\end{align}

\subsubsection{Minorizing the Achievable-Rate Objective}
\label{subsec:rate_minorization}

After eliminating the attitude variable $\boldsymbol{\omega}$, the instantaneous achievable rate at
Bob in slot $n$ is expressed as
\begin{equation}
R_{\rmb}(\mathbf p[n],\mathbf a[n])
\triangleq
\log_2\!\left(
1+\rho_0 \hat F_{\rmb}(\mathbf p[n],\mathbf a[n])
\right),
\label{eq:Rn_def_mm}
\end{equation}
where $\hat F_{\rmb}(\mathbf p[n],\mathbf a[n])$ is the reduced
Rician-averaged channel gain from the UAV to Bob. Accordingly, the average
achievable rate is given by
\begin{equation}
\bar R_{\rmb}(\mathbf P,\mathbf A)
=
\frac{1}{N}
\sum_{n=1}^{N}
R_{\rmb}(\mathbf p[n],\mathbf a[n]).
\label{eq:Rbar_def_mm}
\end{equation}
Since $R_{\rmb}(\mathbf p[n],\mathbf a[n])$ is generally non-concave in the
coupled variables $(\mathbf p[n],\mathbf a[n])$, we construct a local concave
minorizer at each MM iteration. Let
$\{\mathbf p^{(r)}[n],\mathbf a^{(r)}[n]\}_{n=1}^{N}$ denote the current
feasible point, and define $R_{\rmb}^{(r)}[n]
\triangleq
R_{\rmb}
\big(\mathbf p^{(r)}[n],\mathbf a^{(r)}[n]\big)$.
Then, a concave local lower bound of $R_{\rmb}[n]$ is given by
\begin{align}
\!\!\!\!\! R_{\rmb}^{\rm lb}[n]
\!\!=&
R_{\rmb}^{(r)}[n]
+
\big(\nabla_{\mathbf p}R_{\rmb}^{(r)}[n]\big)^T
\big(\mathbf p[n]-\mathbf p^{(r)}[n]\big)
\nonumber\\
&\!+
\big(\nabla_{\mathbf a}R_{\rmb}^{(r)}[n]\big)^T
\big(\mathbf a[n]-\mathbf a^{(r)}[n]\big)
\nonumber\\
&\!\!-\!\!
\frac{\tau_{\rmp}^{(r)}\![n]}{2}\!
\big\|\mathbf p[n]\!\!-\!\mathbf p^{\!(r)}\![n]\big\|^2
\!\!-\!\!
\frac{\tau_{\rma}^{(r)}\![n]}{2}
\big\|\mathbf a[n]\!\!-\!\mathbf a^{\!(r)}[n]\!\big\|^2\!,\!\!\!
\label{eq:R_minorizer}
\end{align}
where $\tau_{\rmp}^{(r)}[n]\ge0$ and $\tau_{\rma}^{(r)}[n]\ge0$ are curvature parameters chosen such that $R_{\rmb}^{\rm lb}[n] \le R_{\rmb}(\mathbf p[n],\mathbf a[n])$ holds within the trust region. 
This bound is tight at the current iterate and
matches the first-order derivatives of $R_{\rmb}[n]$, and
the required gradients are given by
{\small\begin{align}
&\!\!\!\!\!\!\!\nabla_{\!\mathbf p} R_{\rmb}^{(r)}\![n]
\!\!=\!\!
\frac{\rho_0}
{\ln \!2\!\left(\!1\!+\!\rho_0 \hat F_{\rmb}\!
\big(\mathbf p^{(r)}\![n],\mathbf a^{\!(r)}\![n]\big)\!\!\right)}\!
\nabla_{\!\mathbf p}\hat F_{\rmb}\!
\big(\mathbf p^{\!(r)}\![n],\mathbf a^{\!(r)}\![n]\big)\!,\!\!\!
\label{eq:gradp_R_mm}
\\&\!\!\!\!\!\!\!\nabla_{\!\mathbf a} R_{\rmb}^{(r)}\![n]
\!\!=\!\!
\frac{\rho_0}
{\ln \!2\!\left(\!1\!+\!\rho_0 \hat F_{\rmb}\!
\big(\mathbf p^{(r)}\![n],\mathbf a^{\!(r)}\![n]\big)\!\!\right)}\!
\nabla_{\!\mathbf a}\hat F_{\rmb}\!
\big(\mathbf p^{\!(r)}\![n],\mathbf a^{\!(r)}\![n]\big)\!.\!\!\!
\label{eq:grada_R_mm}
\end{align}}Therefore, the concave lower bound of the average rate is
\begin{equation}
{\bar R}_{\rmb}^{\rm lb}
\triangleq
\frac{1}{N}
\sum_{n=1}^{N}
R_{\rmb}^{\rm lb}[n].
\label{eq:Rbar_minorizer}
\end{equation}
The MM update is then obtained by maximizing ${\bar R}_{\rmb}^{\rm lb}$.

\subsubsection{Majorizing the Covert Constraint}
Based on \eqref{covert_measure}, we define the covert metric in slot $n$ as 
\begin{equation}
C(\mathbf p[n],\mathbf a[n])
\triangleq
\rho_0 \hat F_{\rmw}(\mathbf p[n],\mathbf a[n]).
\label{eq:Cn_def_mm} 
\end{equation}
Since $C(\mathbf p[n],\mathbf a[n])$ is generally non-convex, directly
enforcing \eqref{eq:Cn_def_mm} is intractable. To obtain a conservative convex restriction, at iteration $r$, we further define $C^{(r)}[n]
\triangleq
C\big(\mathbf p^{(r)}[n],\mathbf a^{(r)}[n]\big)$. Thus we can construct the following convex local upper bound: 
\begin{align}
\!\!\!\!\! C^{\rm up}[n]
\!\!=&
C^{(r)}[n]
+
\big(\nabla_{\mathbf p}C^{(r)}[n]\big)^T
\big(\mathbf p[n]-\mathbf p^{(r)}[n]\big)
\nonumber\\
&\!\!+
\big(\nabla_{\mathbf a}C^{(r)}[n]\big)^T
\big(\mathbf a[n]-\mathbf a^{(r)}[n]\big)
\nonumber\\
&\!\!+\!\!
\frac{\xi_{\rmp}^{(r)}\![n]}{2}
\big\|\mathbf p[n]\!\!-\!\!\mathbf p^{(r)}\![n]\big\|^2
\!\!+\!\!
\frac{\xi_{\rma}^{(r)}\![n]}{2}
\big\|\mathbf a[n]\!\!-\!\mathbf a^{(r)}\![n]\big\|^2\!,
\label{eq:C_majorizer}\!\!
\end{align}
where $\xi_{\rmp}^{(r)}[n]\ge0$ and $\xi_{\rma}^{(r)}[n]\ge0$ are local curvature parameters selected such that $C(\mathbf p[n],\mathbf a[n]) \le C^{\rm up}[n]$ holds within the trust region. 
The gradients in \eqref{eq:C_majorizer} are given by
\begin{align}
\nabla_{\mathbf p}C^{(r)}[n]
&=
\rho_0
\nabla_{\mathbf p}\hat F_{\rmw}
\big(\mathbf p^{(r)}[n],\mathbf a^{(r)}[n]\big),
\label{eq:gradp_C_mm}
\\
\nabla_{\mathbf a}C^{(r)}[n]
&=
\rho_0
\nabla_{\mathbf a}\hat F_{\rmw}
\big(\mathbf p^{(r)}[n],\mathbf a^{(r)}[n]\big).
\label{eq:grada_C_mm}
\end{align}
Therefore, the conservative surrogate constraint is given by\vspace{-1mm}
\begin{equation}
C^{\rm up}[n]
\le
\sqrt{\frac{2\epsilon}{L}},
\qquad \forall n\in\mathcal N.
\label{eq:C_majorizer_constraint} 
\end{equation}

\subsubsection{Local Convexification of the Angular Smoothness Constraint}
Define the original angular smoothness function as
\begin{equation}
S(\mathbf a[n],\mathbf a[n+1])
\triangleq
\left\|
\frac{\mathbf a[n+1]+\mathbf g}
{\|\mathbf a[n+1]+\mathbf g\|}
-
\frac{\mathbf a[n]+\mathbf g}
{\|\mathbf a[n]+\mathbf g\|}
\right\|^2 .
\end{equation}
At iteration $r$, let $S^{(r)}[n]=S(\mathbf a^{(r)}[n],\mathbf a^{(r)}[n+1])$. We then construct the following convex upper bound:
\begin{align}
\!\!\!\!S^{\rm up}[n]
\!\!=&
S^{(r)}[n]
+
(\nabla_{\mathbf a[n]}S^{(r)}[n])^T
\big(\mathbf a[n]-\mathbf a^{(r)}[n]\big)
\nonumber\\*
&+
(\nabla_{\mathbf a[n+1]}S^{(r)}[n])^T
\big(\mathbf a[n+1]-\mathbf a^{(r)}[n+1]\big)
\\*
&\!\!+\!\!
\frac{\zeta^{(r)}\![n]}{2}
\big\|\mathbf a[n]\!\!-\!\!\mathbf a^{(r)}[n]\big\|^2
\!\!+\!\!
\frac{\zeta^{(r)}\![n]}{2}
\big\|\mathbf a[n\!\!+\!\!1]\!\!-\!\!\mathbf a^{(r)}[n\!\!+\!\!1]\big\|^2 \!\!,\nonumber
\end{align}
where $\zeta^{(r)}[n]\ge 0$ is chosen such that $S(\mathbf a[n],\mathbf a[n+1]) \le S^{\rm up}[n]$
holds within the trust region.
Accordingly, the angular smoothness constraint is conservatively enforced by
\begin{equation}
S^{\rm up}[n]
\le
\delta_{\rm rot},
\qquad
\forall n\in\mathcal N\setminus\{N\}.\label{covert_cons}
\end{equation}

\subsubsection{MM-Based Convex Approximation of Problem (P1)}
To ensure the validity of the local minorizer and majorizer, we impose the following
trust-region constraints:
\begin{align}
\|\mathbf{p}[n]-\mathbf{p}^{(r)}[n]\|
&\le \Delta_\rmp^{(r)}, \qquad \forall n\in\mathcal{N},
\label{eq:trust_p_mm}
\\
\|\mathbf{a}[n]-\mathbf{a}^{(r)}[n]\|
&\le \Delta_\rma^{(r)}, \qquad \forall n\in\mathcal{N},
\label{eq:trust_a_mm}
\end{align}
where $\Delta_\rmp^{(r)}$ and $\Delta_\rma^{(r)}$ are the trust-region radii at iteration $r$.
Therefore, at iteration $r$, the MM-based convex approximation of problem (P1) is given by
\begin{subequations}
\begin{align}
\text{(P1.1):}\quad
\max_{\mathbf{P},\mathbf{V},\mathbf{A}}
\quad &
\frac{1}{N}\sum_{n=1}^{N} R_{\rmb}^{\rm lb}[n]
\label{eq:P2_obj}
\\
\text{s.t.}~\quad
&
\eqref{p0_start_final}-\eqref{p0_cons_vaz}, 
\eqref{p0_cons_sec},
\eqref{p0_cons_a1}-\eqref{p0_cons_a4},\nn\\
&\eqref{eq:C_majorizer_constraint},\eqref{covert_cons},\eqref{eq:trust_p_mm},\eqref{eq:trust_a_mm}.\nn
\end{align}
\end{subequations}
Since the objective \eqref{eq:P2_obj} is concave and all constraints are convex, problem (P1.1) is a convex optimization problem and can be efficiently solved by standard CVX solvers\cite{cvx}.

\subsection{Algorithm Summary and Complexity Analysis}

The proposed MM-based procedure is summarized in Algorithm \ref{alg:MM1}.
We next analyze the computational complexity of Algorithm \ref{alg:MM1}. 
At each MM iteration, the reduced attitude vectors, channel gains, and the gradients required for constructing $R_{\rmb}^{\rm lb}[n]$, $C^{\rm up}[n]$, and $S^{\rm up}[n]$ are evaluated for all time slots. Since these operations are performed independently over $N$ time slots, the corresponding complexity is $\mathcal O(N)$.
The main computational cost comes from solving the convex approximation problem (P1.1). The optimization variables are $\mathbf P$, $\mathbf V$, and $\mathbf A$, which contain approximately $9N$ real-valued variables in total. The numbers of equality constraints, affine inequalities, and second-order-cone representable constraints in (P1.1) all scale linearly with $N$. Therefore, using a standard interior-point solver, the worst-case complexity of solving (P1.1) at each MM iteration is on the order of $\mathcal O\big((9N)^3\big)$.
The additional $\mathcal O(N)$ cost for constructing the surrogate functions is dominated by the convex optimization step.
Let $I_{\rm MM}$ denote the number of MM iterations required for convergence. The overall computational complexity of the proposed algorithm is therefore $\mathcal O\left(I_{\rm MM}\left((9N)^3+N\right)\right)$.

\begin{algorithm}[!t]
\caption{Proposed MM-based algorithm for (P1)}
\label{alg:MM1}
\textbf{Initialize:}{
Initial feasible $\mathbf{P}^{(0)}$, $\mathbf{V}^{(0)}$, $\mathbf{A}^{(0)}$, and $r=0$; 
}

\Repeat{convergence}{
Compute $\boldsymbol{\omega}^{(r)}[n]$, $\mathbf{J}_{\omega}^{(r)}[n]$, and the gradients of $\hat F_{\rmb}$ and $\hat F_{\rmw}$ for all $n\in\mathcal{N}$\;
Construct $R_{\rmb}^{\rm lb}[n]$, \!$C^{\rm up}[n]$, and $S^{\rm up}[n]$ for all $n\!\in\!\mathcal{N}$\;

Solve the convex approximation (P1.1) and obtain 
$\mathbf{P}^{(r+1)}$, $\mathbf{V}^{(r+1)}$, and $\mathbf{A}^{(r+1)}$\;

Set $r\leftarrow r+1$\;
}

\Return{$\mathbf{P}^{\star}=\mathbf{P}^{(r)}$, 
$\mathbf{V}^{\star}=\mathbf{V}^{(r)}$, and 
$\mathbf{A}^{\star}=\mathbf{A}^{(r)}$}.
\end{algorithm}

\section{\scalebox{0.96}[1]{Robust Design With Imperfect Attitude Control}}
\label{sec:imperfect_attitude_control}

In the previous sections, the UAV attitude is assumed to follow the dynamics-consistent mapping induced by translational acceleration. In practice, attitude tracking errors, actuator imperfections, and sensing inaccuracies may cause deviations from the nominal reduced attitude. To capture this effect, we develop a robust design under imperfect attitude control.
\vspace{-2mm}
\subsection{System Model Under Imperfect Attitude Control}

In the ideal model, the UAV body vertical axis is assumed to perfectly follow the dynamics-consistent reduced attitude induced by the translational
acceleration.
In practice, due to attitude tracking errors, actuator imperfections, and sensing inaccuracy, the actual body vertical axis may deviate from ${\boldsymbol{\omega}}[n]$. Let
$\breve{\boldsymbol{\omega}}[n]$ denote the actual body vertical axis. The imperfect reduced attitude is modeled as
\vspace{-1mm}
\begin{equation}
\breve{\boldsymbol{\omega}}[n]
=
\frac{\boldsymbol{\omega}[n]+\mathbf e_{\omega}[n]}
{\|\boldsymbol{\omega}[n]+\mathbf e_{\omega}[n]\|},
\qquad \forall n\in\mathcal N ,
\label{eq:omega_actual_imp}\vspace{-1mm}
\end{equation}
where $\mathbf e_{\omega}[n]$ denotes the attitude-control error. Since
$\boldsymbol{\omega}[n]$ lies on the unit sphere, the perturbation is restricted
to the tangent space of the nominal attitude:\vspace{-1mm}
\begin{equation}
\mathcal E[n]
=
\left\{
\mathbf e_{\omega}[n]:
\boldsymbol{\omega}^T[n]\mathbf e_{\omega}[n]=0,\;
\|\mathbf e_{\omega}[n]\|\le \varepsilon_{\omega}
\right\}.
\label{eq:error_set_imp}\vspace{-1mm}
\end{equation}
For node $\dagger$, the transmitter-side off-axis angle is  
\begin{equation}
\breve\vartheta_{{\rm tx},\dagger}[n]
=
\arccos\!\left(
\frac{\mathbf d_\dagger^T[n]\breve{\boldsymbol{\omega}}[n]}
{d_\dagger[n]}
\right),
\label{eq:theta_tx_actual_imp}
\end{equation}
while the receiver-side angle $\vartheta_{{\rm rx},\dagger}[n]$ remains unchanged
since it is determined by the UAV-node geometry and the fixed ground antenna
orientation. Hence, the actual LoS gain is\vspace{-1mm}
\begin{equation}
\breve{\bar F}_\dagger
\left(
\mathbf p[n],\mathbf a[n],\mathbf e_{\omega}[n]
\right)
=
\frac{
G\!\left(\breve\vartheta_{{\rm tx},\dagger}[n]\right)
G\!\left(\vartheta_{{\rm rx},\dagger}[n]\right)
}
{d_\dagger^2[n]}.
\label{eq:Fbar_actual_imp}\vspace{-1mm}
\end{equation}
Since the averaged NLoS gain is modeled as a distance-dependent residual scattering term, it is not directly affected by the attitude perturbation $\tilde F_\dagger(\mathbf p[n])
=
\tilde{\eta}_{\dagger}[n]d_\dagger^{-\tilde\alpha }[n]$.
Therefore, the actual Rician-averaged gain under attitude error is 
\vspace{-1mm}
\begin{equation}
\breve{F}_\dagger
=
\frac{\kappa_\dagger}{1+\kappa_\dagger}
\breve{\bar F}_\dagger\left(
\mathbf p[n],\mathbf a[n],\mathbf e_{\omega}[n]
\right) 
+
\frac{1}{1+\kappa_\dagger}
\tilde F_\dagger(\mathbf p[n]).
\label{eq:Fhat_actual_imp}\vspace{-1mm}
\end{equation}

Since the exact realization of $\mathbf e_{\omega}[n]$ is unknown, we adopt a
worst-case robust design. Specifically, the worst-case Rician-averaged gain for Bob is given by
\begin{equation}
{\breve{F}}_{\rmb}^{\rm lb}
\left(
\mathbf p[n],\mathbf a[n]
\right)
\triangleq
\min_{\mathbf e_{\omega}[n]\in\mathcal E[n]}
\breve{F}_\rmb
\left(
\mathbf p[n],\mathbf a[n],\mathbf e_{\omega}[n]
\right).
\label{eq:Fb_wc_imp}
\end{equation}
The corresponding worst-case achievable rate is
\begin{equation}
\breve R_{\rmb}(\mathbf p[n],\mathbf a[n])
=
\log_2
\left(
1+\rho_0
{\breve F}_{\rmb}^{\rm lb}
\left(
\mathbf p[n],\mathbf a[n]
\right)
\right).
\label{eq:Rb_wc_imp}
\end{equation}
For Willie, the worst-case Rician-averaged gain is given by
\vspace{-1.5mm}
\begin{equation}
{\breve F}_{\rmw}^{\rm up}
\left( 
\mathbf p[n],\mathbf a[n]
\right)
\triangleq
\max_{\mathbf e_{\omega}[n]\in\mathcal E[n]}
\breve F_{\rmw} 
\left(
\mathbf p[n],\mathbf a[n],\mathbf e_{\omega}[n]
\right).
\label{eq:Fw_wc_imp}\vspace{-1.5mm}
\end{equation}
Accordingly, the robust per-slot covert constraint is given by
\vspace{-1mm}
\begin{equation}
\!\!\!\!\breve C(\mathbf p[n],\mathbf a[n])\triangleq\rho_0
{\breve F}_{\rmw}^{\rm up}
\left(
\mathbf p[n],\mathbf a[n]
\right)
\le
\sqrt{\frac{2\epsilon}{L}},
\forall n\in\mathcal N .
\label{eq:covert_wc_imp}\vspace{-1mm}
\end{equation}
This constraint ensures that the covert requirement is satisfied under the
worst-case attitude perturbation within the uncertainty set
$\mathcal{E}[n]$.

\subsection{Robust Problem Formulation}

Based on the worst-case achievable rate in \eqref{eq:Rb_wc_imp} and the robust covert constraint in \eqref{eq:covert_wc_imp}, the robust trajectory optimization problem under imperfect attitude control is formulated as
\begin{subequations}
\begin{align}
\text{(P2):}\quad
\max_{\mathbf P,\mathbf V,\mathbf A}
\quad&
\frac{1}{N}\sum_{n=1}^{N}
\breve R_{\rmb}(\mathbf p[n],\mathbf a[n])
\label{eq:robust_obj}
\\
\text{s.t.}~\quad
&
\breve C(\mathbf p[n],\mathbf a[n])
\le
\sqrt{\frac{2\epsilon}{L}},
\quad \forall n\in\mathcal N,
\label{eq:robust_covert}
\\
&
S(\mathbf a[n],\mathbf a[n\!+\!1])
\!\!\le\!
\delta_{\rm rot}, \forall n\!\in\!\mathcal N\!\setminus\!\{N\},
\label{eq:robust_smooth}
\\
&
\eqref{p0_start_final}-\eqref{p0_cons_vaz},
\eqref{p0_cons_sec},
\eqref{p0_cons_a1}-\eqref{p0_cons_a4}.
\nonumber
\end{align}
\end{subequations}
Problem (P2) is non-convex due to the worst-case achievable-rate objective, the robust covert constraint, and the nonlinear angular-smoothness constraint. In particular, both $\breve R_{\rmb}(\mathbf p[n],\mathbf a[n])$ and $\breve C(\mathbf p[n],\mathbf a[n])$ depend on the UAV position and acceleration through the robust Rician-averaged gains and the acceleration-induced attitude mapping. Therefore, (P2) is difficult to solve directly. We next develop an MM-based algorithm to obtain a tractable solution.

\subsection{Robust MM-Based Solution}

The robust design involves the worst-case Rician-averaged gains
${\breve F}_{\rmb}^{\rm lb}$ and ${\breve F}_{\rmw}^{\rm up}$, which are
defined over the attitude-error set $\mathcal E[n]$. Directly optimizing over
all possible attitude perturbations is generally intractable. We therefore
derive tractable conservative bounds based on the induced angular uncertainty.
Since $\mathbf e_{\omega}[n]$ lies in the tangent space of
$\boldsymbol{\omega}[n]$ and satisfies
$\|\mathbf e_{\omega}[n]\|\le\varepsilon_{\omega}$, the angular deviation
between the nominal body axis $\boldsymbol{\omega}[n]$ and the actual body
axis $\breve{\boldsymbol{\omega}}[n]$ is upper-bounded by
\vspace{-1mm}
\begin{equation}
\Delta_{\omega}
\triangleq
\arctan(\varepsilon_{\omega}).
\label{eq:delta_omega_bound}\vspace{-1mm}
\end{equation}
Hence, the actual transmitter-side off-axis angle satisfies
\vspace{-1mm}
\begin{equation}
\!\!\!\!\breve\vartheta_{{\rm tx},\dagger}[n]
\in
\left[
\vartheta_{{\rm tx},\dagger}[n]-\Delta_{\omega},
\vartheta_{{\rm tx},\dagger}[n]+\Delta_{\omega}
\right] .\vspace{-1mm}
\label{eq:theta_uncertainty_bound}
\end{equation}

Since the antenna gain $G(\vartheta)$ is non-increasing with respect to the
off-boresight angle, the LoS gain at Bob is conservatively lower-bounded by
\begin{equation}
\breve{\bar F}_{\rmb}^{\rm lb}[n]
=
\frac{
G\!\left(
\min\{\pi,\vartheta_{{\rm tx},\rmb}[n]+\Delta_{\omega}\}
\right)
G\!\left(\vartheta_{{\rm rx},\rmb}[n]\right)
}
{d_{\rmb}^2[n]}.
\label{eq:Fbar_bob_lb}
\end{equation}
Similarly, the LoS gain at Willie is conservatively upper-bounded by
\begin{equation}
\breve{\bar F}_{\rmw}^{\rm up}[n]
\!\!=\!\!
\frac{
G\!\left(
\max\{0,\vartheta_{{\rm tx},\rmw}[n]-\Delta_{\omega}\}
\right)
G\!\left(\vartheta_{{\rm rx},\rmw}[n]\right)
}
{d_{\rmw}^2[n]}.
\label{eq:Fbar_willie_up}
\end{equation}

The averaged NLoS component is retained in the robust model. Since its explicit
attitude dependence has been absorbed into the scenario-level coefficient
$\tilde\eta[n]$, it remains $\tilde F_\dagger[n]
\!=\!
\tilde\eta[n] d_\dagger^{-\tilde\alpha}[n]$.
Therefore, the robust lower bound of Bob's Rician-averaged gain and the robust
upper bound of Willie's Rician-averaged gain are given by
\begin{align}
{\breve F}_{\rmb}^{\rm lb}[n]
&\triangleq
\frac{\kappa_\rmb}{1+\kappa_\rmb}
{\breve{\bar F}}_{\rmb}^{\rm lb}[n]
+
\frac{1}{1+\kappa_\rmb}
\tilde F_{\rmb}[n],
\label{eq:Fb_lb_robust}
\\
{\breve F}_{\rmw}^{\rm up}[n]
&\triangleq
\frac{\kappa_\rmw}{1+\kappa_\rmw}
{\breve{\bar F}}_{\rmw}^{\rm up}[n]
+
\frac{1}{1+\kappa_\rmw}
\tilde F_{\rmw}[n].
\label{eq:Fw_up_robust}
\end{align}
Here, ${\breve F}_{\rmb}^{\rm lb}[n]$ characterizes the worst-case degradation
of Bob's effective channel, while ${\breve F}_{\rmw}^{\rm up}[n]$ characterizes the worst-case enhancement of Willie's effective channel.

Note that the robust reduced problem remains non-convex because
$\breve R_{\rmb}$ is generally non-concave and
$\breve C$ is generally non-convex in the coupled variables
$(\mathbf p[n],\mathbf a[n])$. We therefore apply the same MM principle as in
the nominal-attitude case.
At iteration $r$, let
$\{\mathbf p^{(r)}[n],\mathbf a^{(r)}[n]\}_{n=1}^{N}$
denote the current feasible point, and define $\breve R_{\rmb}^{(r)}[n]
\triangleq
\breve R_{\rmb}
\left(
\mathbf p^{(r)}[n],
\mathbf a^{(r)}[n]
\right)$, $\breve C^{(r)}[n]
\triangleq
\breve C
\left(
\mathbf p^{(r)}[n],
\mathbf a^{(r)}[n]
\right)$.
The required gradients can be obtained by the chain rule from
${\breve F}_{\rmb}^{\rm lb}$ and ${\breve F}_{\rmw}^{\rm up}$.
For the robust objective, we construct the following concave local lower bound:
\begin{align}
\!\!\!\!\! \breve R_{\rmb}^{{\rm lb}}[n]
\!=&
\breve R_{\rmb}^{(r)}[n]
+
\left(
\nabla_{\mathbf p}
{\breve R}_{\rmb}^{(r)}[n]
\right)^T
\left(
\mathbf p[n]-\mathbf p^{(r)}[n]
\right)
\nonumber\\*
&+
\left(
\nabla_{\mathbf a}
{\breve R}_{\rmb}^{(r)}[n]
\right)^T
\left(
\mathbf a[n]-\mathbf a^{(r)}[n]
\right)
\nonumber\\*
&\!\!-\!\!
\frac{\breve\tau_{\rmp}^{(r)}\![n]}{2}
\!\!\left\|
\mathbf p[n]\!\!-\!\!\mathbf p^{(r)}\![n]
\right\|^2\!\!
-\!\!
\frac{\breve\tau_{\rma}^{(r)}\![n]}{2}\!\!
\left\|
\mathbf a[n]\!\!-\!\!\mathbf a^{(r)}[n]
\right\|^2\!\!,
\label{eq:Rlb_minorizer_imp}
\end{align}
where $\breve\tau_{\rmp}^{(r)}[n]\ge0$ and
$\breve\tau_{\rma}^{(r)}[n]\ge0$ are chosen such that $\breve R_{\rmb}^{{\rm lb}}[n]
\le
\breve R_{\rmb} \big(
\mathbf p[n],
\mathbf a[n]
\big)$
holds within the trust region.

Similarly, for the robust covert metric, we construct the following convex
local upper bound:\vspace{-1.5mm}
\begin{align}
\!\!\!\!\! \breve C^{{\rm up}}[n]
\!\!=&
\breve C^{(r)}[n]
+
\left(
\nabla_{\mathbf p}
\breve C^{(r)}[n]
\right)^T
\left(
\mathbf p[n]-\mathbf p^{(r)}[n]
\right)\nn\\
&+
\left(
\nabla_{\mathbf a}
\breve C^{(r)}[n]
\right)^T
\left(
\mathbf a[n]-\mathbf a^{(r)}[n]
\right)\nn\\
&\!\!+\!\!
\frac{\breve\xi_{\rmp}^{(r)}\![n]}{2}\!\!
\left\|
\mathbf p[n]\!\!-\!\!\mathbf p^{(r)}\![n]
\right\|^2
\!\!+\!\!
\frac{\breve\xi_{\rma}^{(r)}\![n]}{2}
\left\|
\mathbf a[n]\!\!-\!\!\mathbf a^{(r)}\![n]
\right\|^2\!\!,\!\!\!
\label{eq:Cup_majorizer_imp}
\end{align} 
where $\breve\xi_{\rmp}^{(r)}[n]\ge0$ and
$\breve\xi_{\rma}^{(r)}[n]\ge0$ are selected such that $ \breve C \left(
\mathbf p[n],
\mathbf a[n]
\right)
\le
\breve C^{{\rm up}}[n]$
holds within the trust region.

Note that the angular smoothness constraint and the trust-region constraints are handled
in the same way as in the nominal-attitude case. Therefore, at iteration $r$, the robust MM-based convex approximation is given by
\vspace{-1mm}\begin{subequations}
\begin{align}
\text{(P2.1):}\quad
\max_{\mathbf P,\mathbf V,\mathbf A}
\quad&
\frac{1}{N}
\sum_{n=1}^{N}
\breve R_{\rmb}^{{\rm lb}}[n]
\label{eq:P3_obj}
\\*
\text{s.t.}~\quad
&
\breve C^{{\rm up}}[n]
\le
\sqrt{\frac{2\epsilon}{L}},
\quad
\forall n\in\mathcal N,
\label{eq:P3_covert}
\\*
&
\eqref{p0_start_final}-\eqref{p0_cons_high},
\eqref{p0_cons_sec},
\eqref{p0_cons_a1}-\eqref{p0_cons_a4},
\nonumber\\*
&
\eqref{covert_cons},
\eqref{eq:trust_p_mm},\eqref{eq:trust_a_mm}.\nn
\end{align} 
\end{subequations}
With properly chosen curvature parameters, the constructed surrogates provide valid local lower and upper bounds within the trust region. Thus, the robust MM approximation is convex and can be efficiently solved by standard convex solvers\cite{cvx}.

\begin{remark}
The robust formulation above is based on a worst-case
approximation of the attitude-control error. Compared with the ideal-attitude
design, it accounts for the degradation of Bob's channel and the
enhancement of Willie's channel caused by bounded attitude perturbations.
Consequently, the resulting trajectory is generally more conservative but more
robust to practical attitude-tracking imperfections.
\end{remark}
\vspace{-5mm}
\subsection{Algorithm Summary and Complexity Analysis}

The proposed robust MM-based procedure is summarized in Algorithm \ref{alg:MM_robust}.
We next analyze its computational complexity. At each MM iteration, the robust LoS bounds, robust Rician-averaged gains, and the gradients required for constructing $\breve R_{\rmb}^{\rm lb}[n]$, $\breve C^{\rm up}[n]$, and $S^{\rm up}[n]$ are evaluated for all time slots. Since these operations are performed independently over $N$ time slots, the corresponding complexity is $\mathcal O(N)$.
The main computational cost comes from solving the robust convex approximation problem (P2.1). The optimization variables are $\mathbf P$, $\mathbf V$, and $\mathbf A$, which contain approximately $9N$ real-valued variables in total. The numbers
of equality constraints, affine inequalities, and second-order-cone representable constraints in (P2.1) scale linearly with $N$. Therefore, using a standard interior-point solver, the worst-case complexity of solving (P2.1) at each MM iteration is on the order of $\mathcal O((9N)^3)$. Let $I_{\rm RMM}$ denote the number of robust MM iterations required for convergence. The overall computational complexity is therefore $\mathcal O\left(I_{\rm RMM}\left((9N)^3+N\right)\right)$.
Compared with the nominal case, the robust design has the same asymptotic complexity, since it optimizes the same variables and preserves the same order of constraints. The additional robust gain and gradient evaluations only increase the per-iteration constant factor.
\begin{algorithm}[!t]
\caption{\!Proposed MM-based algorithm for (P2)}
\label{alg:MM_robust}
\textbf{Initialize:}{
Initial feasible $\mathbf{P}^{(0)}$, $\mathbf{V}^{(0)}$, $\mathbf{A}^{(0)}$, and $r=0$;
}

\Repeat{convergence}{
Compute $\boldsymbol{\omega}^{(r)}[n]$, $\Delta_{\omega}$, the robust LoS bounds 
$\breve{\bar F}_{\rmb}^{\rm lb}[n]$ and $\breve{\bar F}_{\rmw}^{\rm up}[n]$ for all $n\in\mathcal{N}$\;

Compute the gradients of $\!\breve F_{\rmb}^{\rm lb}\!$ and $\!\breve F_{\rmw}^{\rm up}\!$ for all $n\!\in\!\mathcal{N}$\;

Construct $\breve R_{\rmb}^{\rm lb}[n]$, $\!\breve C^{\rm up}[n]$, and $\!S^{\rm up}[n]$ for all $n\!\in\!\mathcal{N}$\;

Solve the robust convex approximation  (P2.1)  and obtain  $\mathbf{P}^{(r+1)}$, $\mathbf{V}^{(r+1)}$, and $\mathbf{A}^{(r+1)}$\;

Set $r\leftarrow r+1$\;
}

\Return{$\mathbf{P}^{\star}=\mathbf{P}^{(r)}$, 
$\mathbf{V}^{\star}=\mathbf{V}^{(r)}$, and 
$\mathbf{A}^{\star}=\mathbf{A}^{(r)}$}.
\end{algorithm}

\section{Numerical Results}
\label{sec:simulation}
 In this section, we evaluate the proposed framework by validating the KL-divergence approximation and MM convergence, quantifying the performance gain over the conventional point-mass design, and examining the robustness against attitude-control errors. Unless otherwise specified, the simulation parameters are summarized in Table~\ref{tab:simulation_parameters}.
 The selected mission and flight parameters ensure physically feasible UAV maneuvering, with the prescribed mobility limits allowing the UAV to complete the mission while satisfying the thrust constraint. The antenna parameters follow the 3GPP TR 38.901 single-element radiation pattern~\cite[Table 7.3-1]{3gpp38901}. Since the antenna gain in \eqref{eq:antenna_gain_pattern} is peak-normalized, the peak gain $G_0=10^{G_{\rmE,\max}/10}$ is absorbed into the reference SNR $\rho_0$. For the robust design under imperfect attitude control, two attitude-axis error bounds are considered: $\epsilon_{\omega}\!=\!0.1$ and $\epsilon_{\omega}\!=\!0.05$, corresponding to maximum angular misalignments of $\arctan(0.1)\!\approx\! 5.71^{\circ}$ and $\arctan(0.05)\approx 2.86^{\circ}$, respectively, enabling robustness evaluation under different attitude-control accuracies.
To assess the practical validity of the conventional point-mass assumption, we consider the following two benchmarks.
\begin{itemize}
    \item \textbf{Point-Mass Trajectory Design:} The UAV trajectory is optimized and evaluated using the conventional point-mass model that ignores attitude variations.
    \item \textbf{Point-Mass Actual-Attitude:} The point-mass optimized trajectory is retained, with its performance re-evaluated using the actual acceleration-induced UAV attitude.
\end{itemize}
Based on the above parameter settings and benchmark design, we further evaluate both the perfect-attitude-control and robust imperfect-attitude-control cases, thereby quantifying the gains brought by dynamics consistency and robustness-aware design.

\begin{table}[!t]
\centering\vspace{-2mm}
\caption{Simulation Parameters}\vspace{-2mm}
\label{tab:simulation_parameters}
\footnotesize
\renewcommand{\arraystretch}{1}
\begin{tabular}{@{}p{0.43\columnwidth}p{0.51\columnwidth}@{}}
\toprule
\textbf{Parameter} & \textbf{Value} \\
\midrule
Initial UAV location
& $\mathbf p_{\rm I}=[0,0,80]^T$ m, 
 \\
Final UAV location &$\mathbf p_{\rm F}=[200,200,80]^T$ m
\\

Bob location
& $\mathbf p_{\rmb}=[160,60,0]^T$ m\\
 Willie location
& 
$\mathbf p_{\rmw}=[135,95,0]^T$ m \\
Mission duration
& $T=20$ s \\

Altitude range
& $50\le z[n]\le130$ m\cite{He-CovertUAV-2026} \\

Speed/acceleration limits
& $V_{\max}=20$ m/s, $A_{\max}=5$ m/s$^2$ \\

Mass 
& $m_\rmu=1$ kg \\
Gravity 
& $g=9.81$ m/s$^2$  \\
Thrust
&  $T_{\max}=20$ N \\
Roll/pitch limits
& $|\phi[n]|\le40^\circ$, $|\theta[n]|\le40^\circ$ \\

Attitude smoothness
& $\delta_{\rm rot}=0.20$ \\

Transmit/noise power
& $P_{\rmt}=1$ W, $\sigma^2=-90$ dBm\cite{xu2024secret} \\

Half-power beamwidth
& $\vartheta_{3{\rm dB}}=65^\circ$ ~\cite{3gpp38901} \\
Maximum side-lobe attenuation
& ${\rm SLA}=30$ dB ~\cite{3gpp38901} \\
Peak antenna gain
& 
$G_{\rm E,\max}=8$ dBi~\cite{3gpp38901} \\
Rician factors
& $\kappa_{\rmb}=\kappa_{\rmw}=9$~\cite{qiao_active_attitude_uav} \\

NLoS parameters
& $\tilde\alpha=3$, $\tilde{\eta}_\rmb[n]=\tilde{\eta}_\rmw[n]=0.3$~\cite{qiao_active_attitude_uav} \\

Covertness parameters
& $\epsilon=0.01$, $L=10$ \\

Attitude-error bounds
& $\epsilon_\omega\in\{0.05,0.1\}$ \\
\bottomrule
\end{tabular}
\end{table}

\begin{figure}[t]
\centering\vspace{-4mm}
\includegraphics[width=1\linewidth]{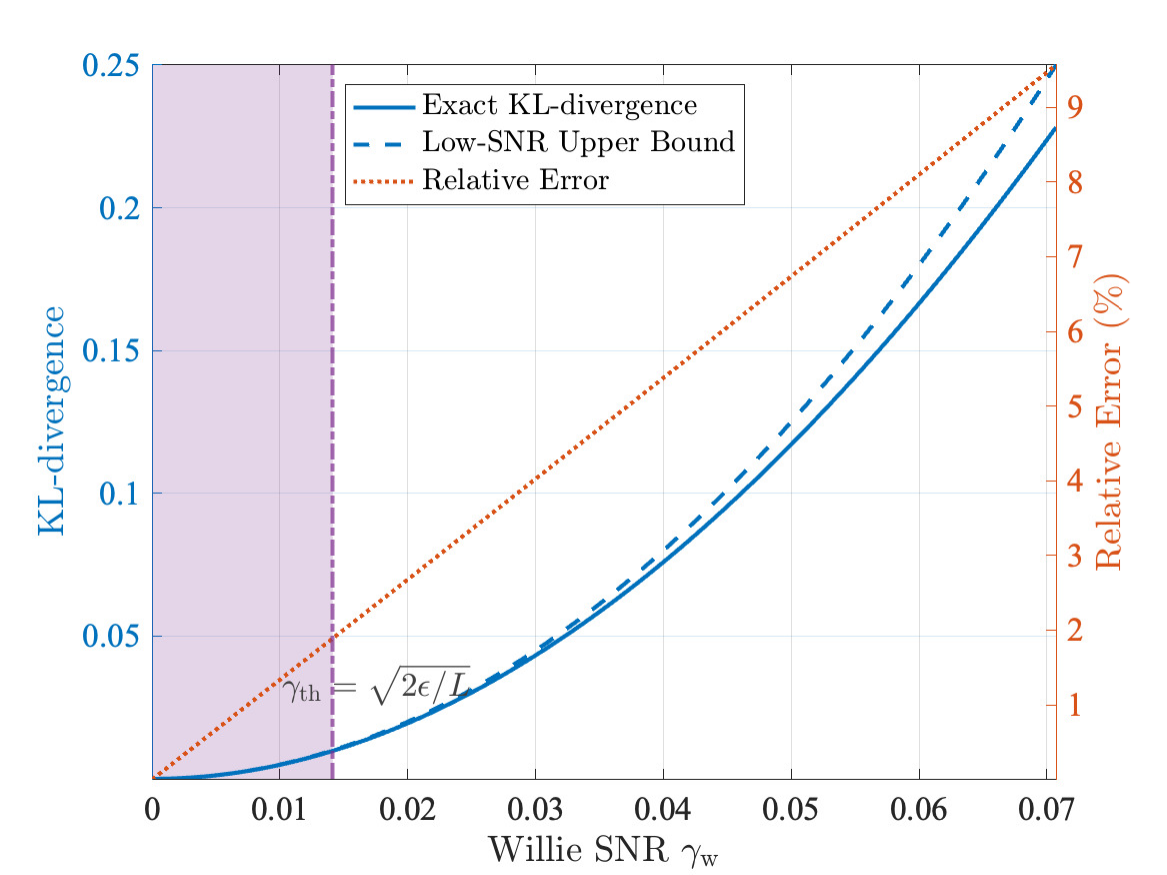}\vspace{-2mm}
\caption{Accuracy of the KL divergence approximation.}\vspace{-1mm}
\label{fig:Fig_KL_Approximation}
\end{figure}

To validate the accuracy of the adopted covert-constraint approximation, Fig.~\ref{fig:Fig_KL_Approximation} compares the exact KL divergence with the adopted low-SNR quadratic upper bound. The shaded region corresponds to the operating regime enforced by the covert constraint, i.e., $\gamma_{\rm w}\le \gamma_{\rm th}=\sqrt{2\epsilon/L}$. As shown in Fig.~\ref{fig:Fig_KL_Approximation}, the two curves are nearly indistinguishable within this region, and the relative approximation error remains small. In addition, the quadratic expression slightly upper-bounds the exact KL divergence, which makes the resulting covert constraint conservative. This confirms that the proposed formulation preserves the KL-divergence-based covertness requirement while enabling a tractable optimization design.

\begin{figure}
\centering\vspace{-4mm}
\includegraphics[width=1\linewidth]{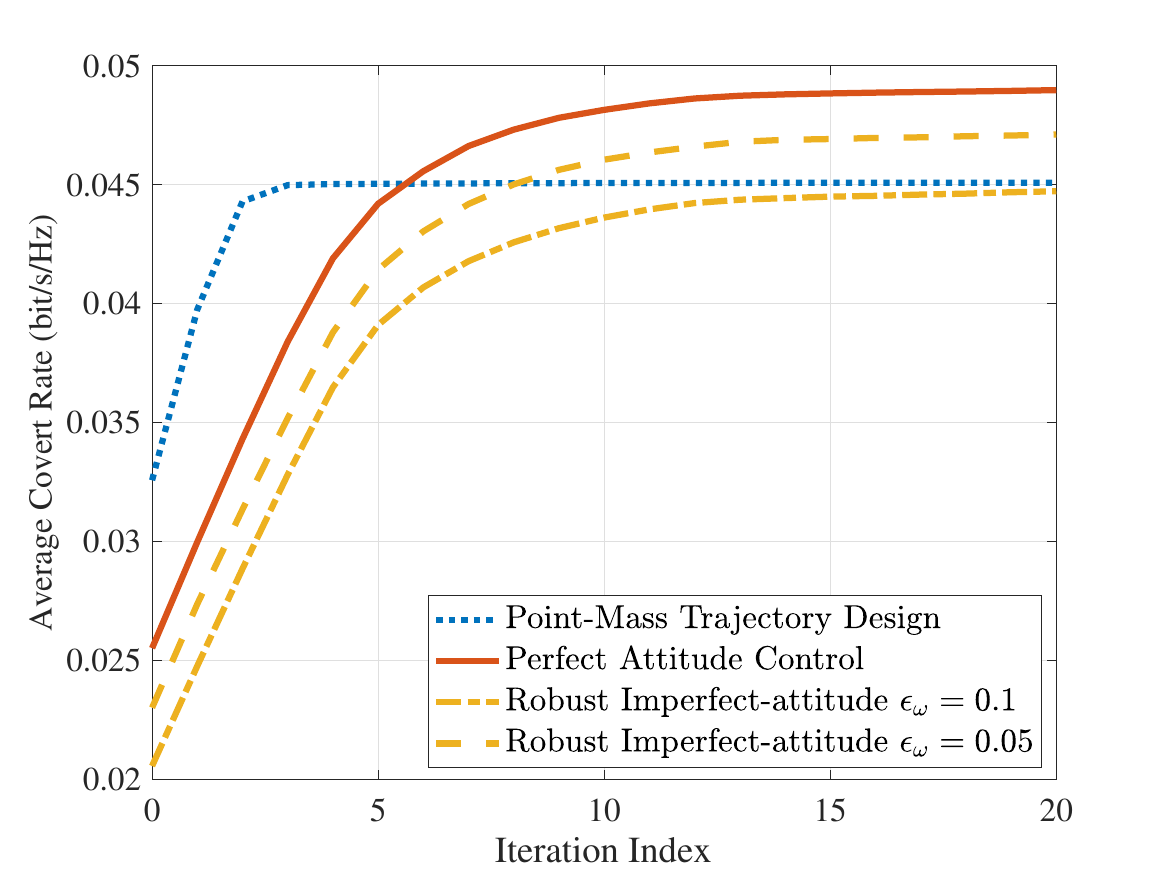}\vspace{-2mm}
\caption{Convergence behavior of the conventional point-mass design and the proposed MM-based algorithms.}
\label{fig:Fig_Convergence}
\end{figure}

Fig.~\ref{fig:Fig_Convergence} shows the convergence of the conventional point-mass design, the nominal MM-based design in Algorithm~\ref{alg:MM1}, and the robust MM-based design in Algorithm~\ref{alg:MM_robust} with $\epsilon_\omega=0.05$ and $0.1$. Note that all curves stabilize within a limited number of iterations. The point-mass design converges faster due to its simpler attitude-agnostic structure, but achieves a lower converged rate than Algorithm~\ref{alg:MM1} and the robust design with $\epsilon_\omega=0.05$.
After convergence, Algorithm~\ref{alg:MM1} achieves the highest average covert rate, while Algorithm~\ref{alg:MM_robust} sacrifices some performance to ensure robustness against attitude-control errors. Even with a relatively large error bound of $\epsilon_\omega=0.1$, the robust design maintains a rate comparable to the idealized performance predicted by the conventional point-mass model, demonstrating a favorable performance--robustness tradeoff.

\begin{figure}[t]
\centering\vspace{-4mm}
\includegraphics[width=1\linewidth]{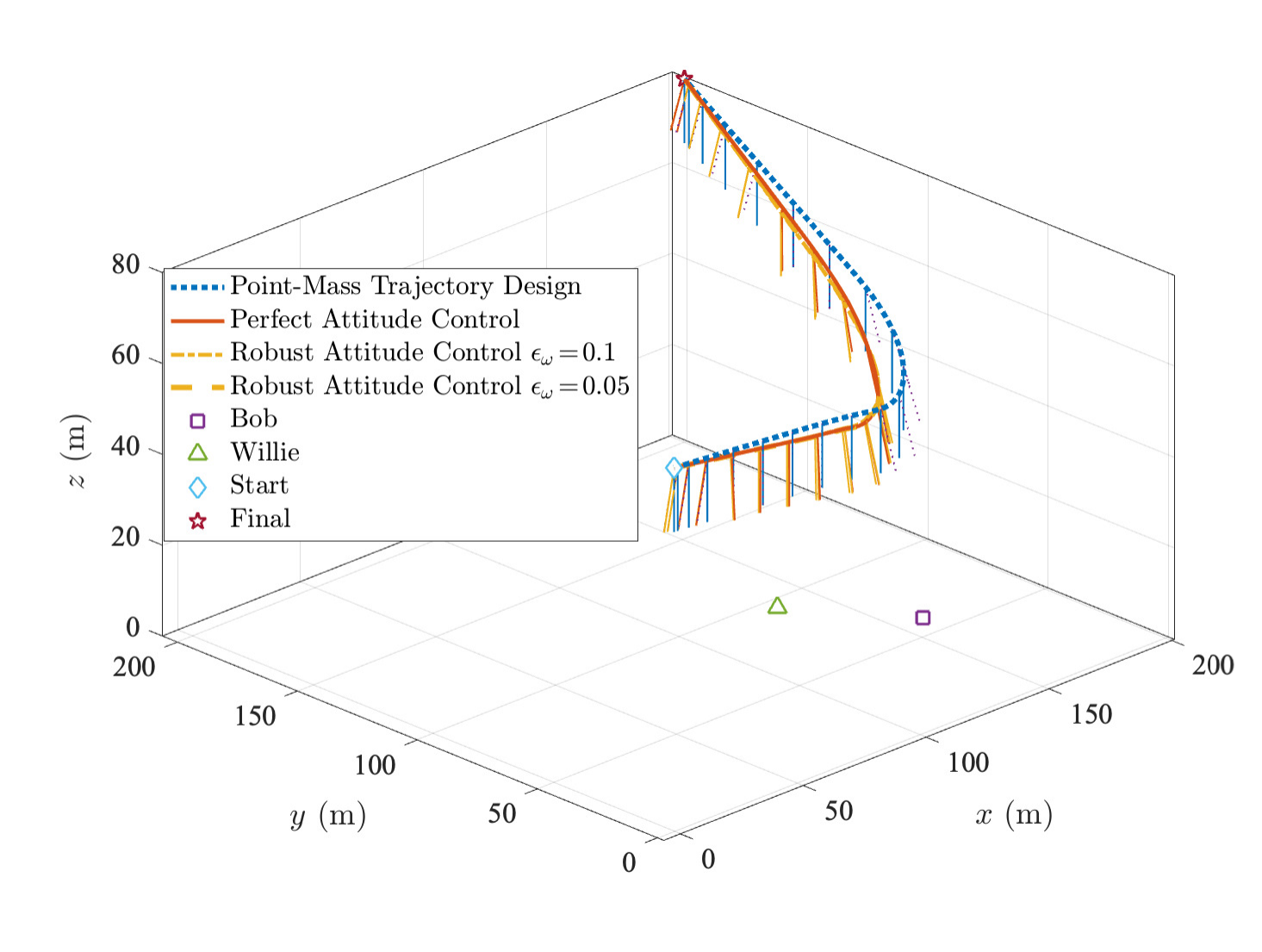}\vspace{-4mm}
\caption{Optimized UAV trajectories and antenna boresight directions under different schemes.}
\vspace{-1mm}
\label{fig:Fig_3D_Trajectory}
\end{figure}

Fig.~\ref{fig:Fig_3D_Trajectory} illustrates the optimized UAV trajectories and antenna boresight directions. The proposed perfect attitude-control design differs from the conventional point-mass trajectory design because it explicitly incorporates the dynamics-consistent acceleration-attitude coupling. By adjusting the acceleration-induced attitude, the UAV can steer its directional antenna more effectively toward Bob while avoiding excessive exposure toward Willie. In contrast, the point-mass design ignores the antenna-orientation effect caused by UAV attitude variation and therefore cannot actively coordinate trajectory and antenna pointing. For the same point-mass trajectory, the purple dotted boresight directions represent the Point-mass Actual-Attitude case obtained by applying the acceleration-attitude mapping to the realized motion. The corresponding practical covert communication performance is evaluated later.
The robust imperfect-attitude design exhibits a more conservative trajectory and boresight pattern, since it accounts for worst-case attitude deviations. Even so, it maintains the key behavior of directing the antenna toward the legitimate receiver while limiting the received power at the warden. This confirms that dynamics-consistent attitude control provides an additional DoF for improving covert communication rate.

\begin{figure}[!t]
\centering\vspace{-3mm}
\includegraphics[width=1\linewidth]{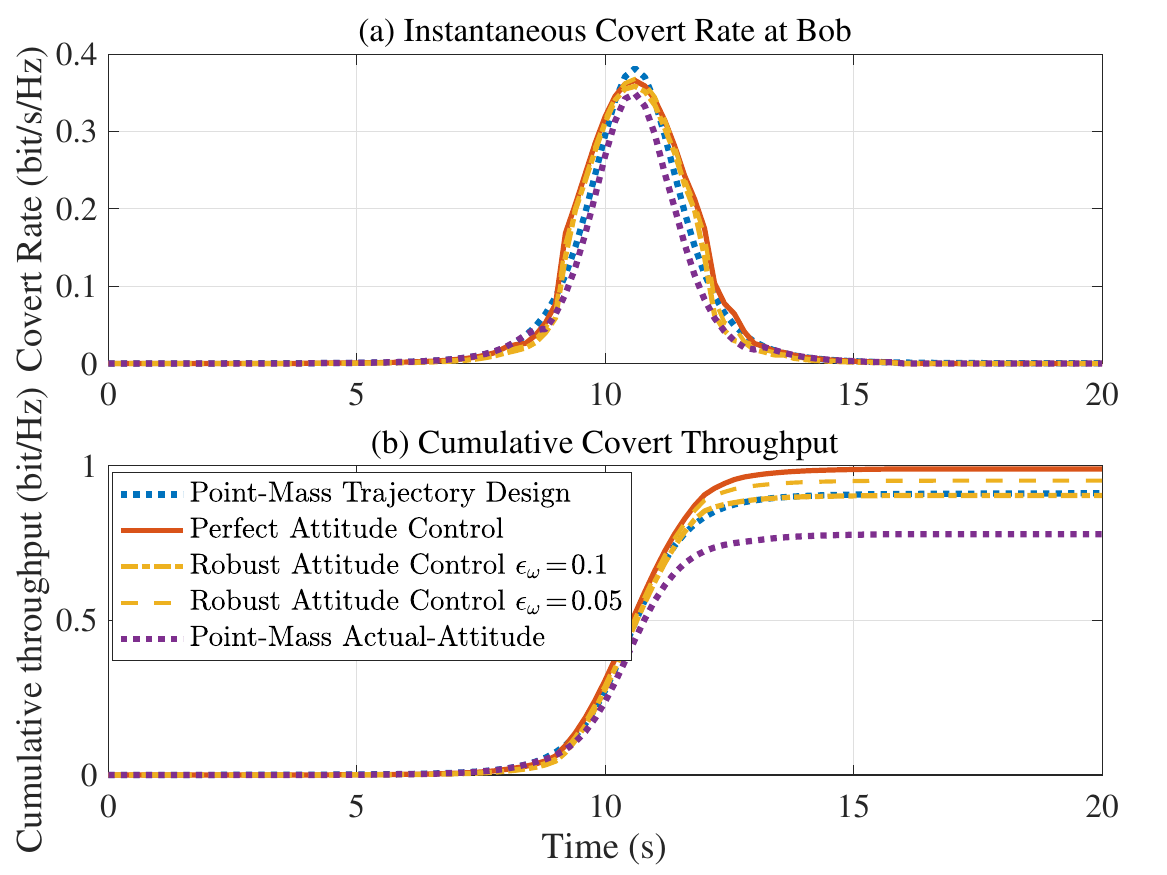}\vspace{-2mm}
\caption{Instantaneous achievable covert rate and cumulative covert throughput at Bob under different schemes.}
\label{fig:Fig_Rate_Throughput}
\end{figure}

Fig.~\ref{fig:Fig_Rate_Throughput} shows the instantaneous covert rate and cumulative covert throughput at Bob. The Point-mass Trajectory Design curve represents the performance predicted by the conventional point-mass model, whereas the Point-mass Actual-Attitude curve re-evaluates the same optimized trajectory under the actual acceleration-induced UAV attitude. A clear performance gap is observed between the two curves, indicating that neglecting attitude variations leads to an optimistic estimate of the practically achievable covert communication performance. This mismatch highlights the necessity of incorporating dynamics-consistent UAV attitude into trajectory design.
The proposed perfect attitude-control design achieves the highest cumulative covert throughput by jointly exploiting UAV trajectory and acceleration-induced attitude to enhance the desired link while limiting the received power at Willie. The robust imperfect-attitude designs yield slightly lower throughput due to worst-case attitude-error protection, but still clearly outperform the Point-mass Actual-Attitude benchmark. These results demonstrate the effectiveness of the proposed dynamics-consistent trajectory-attitude co-design under both perfect and imperfect attitude control.

\begin{figure}[!t]
\centering\vspace{-4mm}
\includegraphics[width=1\linewidth]{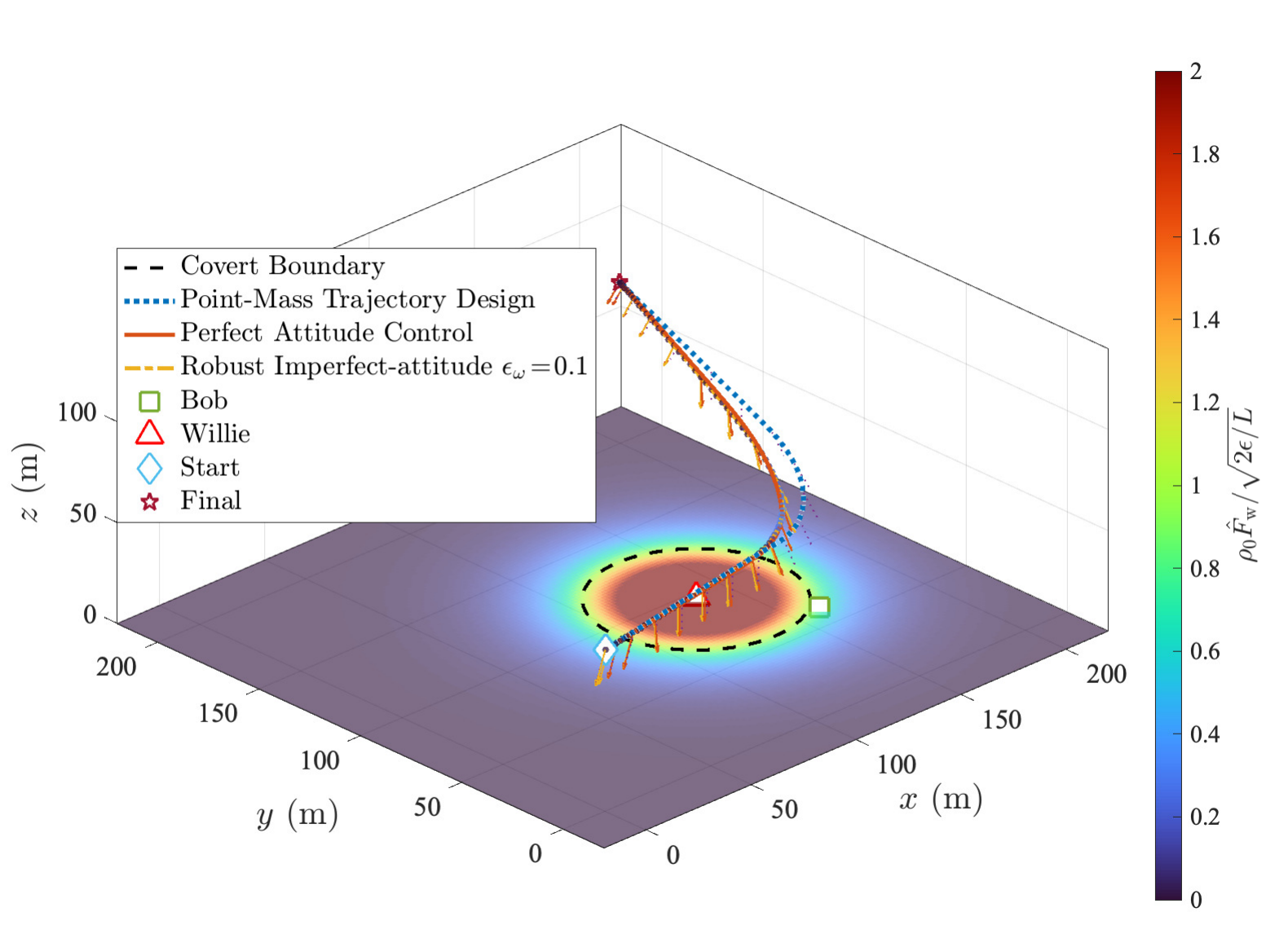}\vspace{-4mm}
\caption{Optimized UAV trajectories over the Willie-side covert-risk map.}
\label{fig:Fig_Covert_Risk}\vspace{-2mm}
\end{figure}
\begin{figure}[!t]
\centering\vspace{-2mm}
\includegraphics[width=1\linewidth]{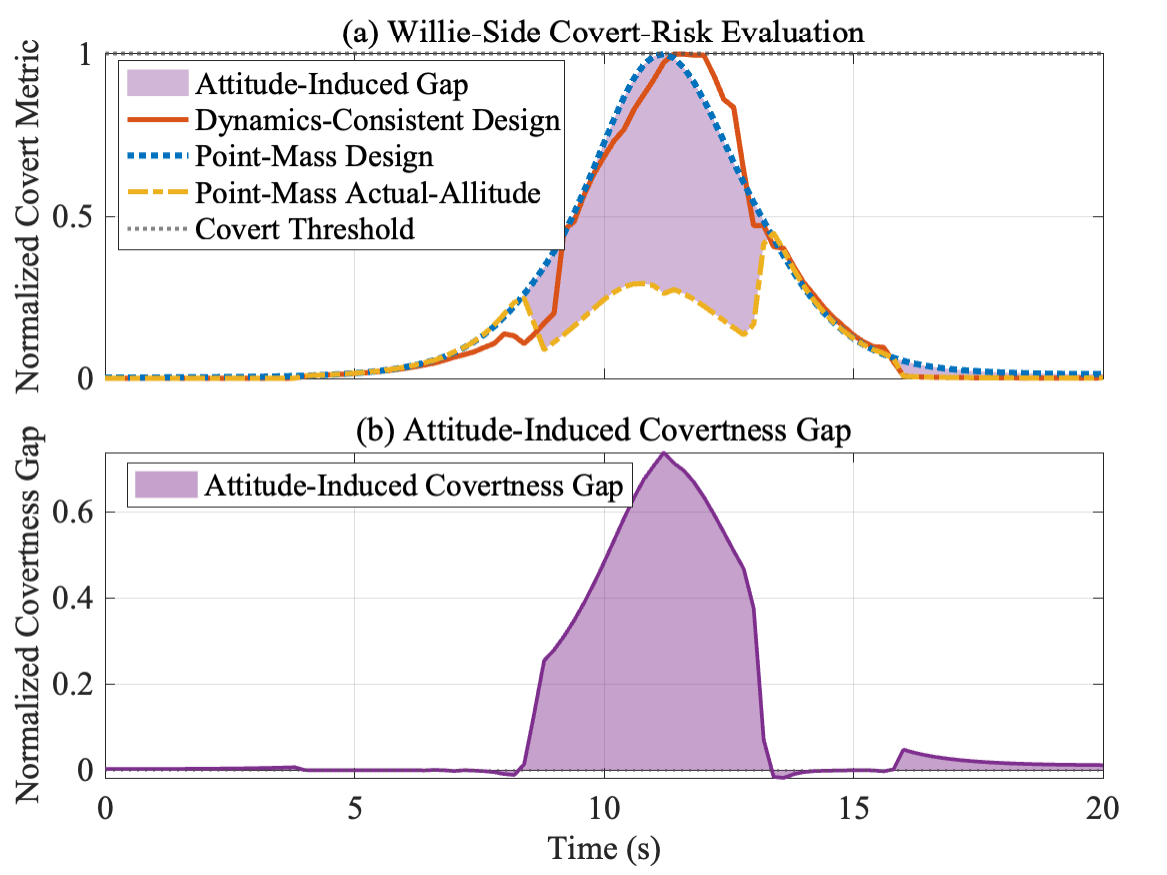}\vspace{-3mm}
\caption{Impact of dynamics-consistent attitude on covert-risk evaluation.}
\label{fig:Fig_Covert_Risk_Mismatch}
\end{figure}
Fig.~\ref{fig:Fig_Covert_Risk} provides a covert-risk-aware visualization of the optimized UAV trajectories. The ground color map represents the normalized Willie-side covert metric $\rho_0\hat F_{\rm w}/\sqrt{2\epsilon/L}$ under a reference altitude and downward antenna orientation. Note that $\rho_0\hat F_{\rm w}/\sqrt{2\epsilon/L}\le1$ indicates that the covert constraint can be satisfied, whereas $\rho_0\hat F_{\rm w}/\sqrt{2\epsilon/L}>1$ implies that Willie’s received power exceeds the allowable covert threshold. Therefore, the black dashed contour, where the normalized metric equals one, can be interpreted as the covert-risk boundary.
The region inside this boundary represents a high-risk area for conventional point-mass trajectory design with a downward-pointing antenna. Since such a design ignores attitude-dependent antenna reorientation, flying above this region may lead to excessive received power at Willie and hence violate the covert constraint. In contrast, the proposed trajectory-attitude design can still pass near this high-risk region by adjusting the acceleration-induced attitude and steering the antenna boresight away from Willie. This reduces the effective antenna gain toward the warden while maintaining a favorable link toward Bob, demonstrating the flexibility provided by the proposed trajectory-attitude co-design.

Fig.~\ref{fig:Fig_Covert_Risk_Mismatch} further illustrates the impact of acceleration-induced attitude variation on Willie-side covert-risk evaluation. In Fig.~\ref{fig:Fig_Covert_Risk_Mismatch} (a), the dynamics-consistent design operates close to the covert threshold during the critical transmission interval, indicating that the proposed design can accurately exploit the available covert margin. For the conventional point-mass design, however, a clear discrepancy is observed between the covert metric predicted by its fixed-orientation model and that obtained by re-evaluating the same trajectory under the actual acceleration-induced attitude. This discrepancy confirms that the point-mass abstraction cannot accurately characterize the Willie-side received power when UAV maneuvers alter directional-antenna orientation.  In Fig.~\ref{fig:Fig_Covert_Risk_Mismatch} (b), we quantify the discrepancy as the normalized attitude-induced covertness gap. In the considered setup, the actual attitude reduces the Willie-side covert metric compared with the point-mass prediction, thereby creating additional covertness margin. These results verify that UAV attitude is not merely a passive flight variable, but an important design factor that reshapes the covert communication constraint through antenna-orientation effects.

\section{Conclusions}
\label{sec:conclusion} 
In this paper, we studied dynamics-consistent trajectory-attitude co-design for UAV covert communications. By modeling the acceleration-induced UAV attitude and its impact on directional channel gains, we formulated a covert transmission rate maximization problem under covertness, UAV motion, thrust, attitude-smoothness, and safety constraints. To solve the resulting non-convex optimization problem, we proposed an MM-based algorithm. Furthermore, we developed a robust extension to account for imperfect attitude control. Extensive numerical simulation results demonstrated that our proposed trajectory design achieves higher covert transmission rate than conventional impractical point-mass trajectory designs.   

We next discuss future research directions. Firstly, our study focused on rotary-wing UAVs, where the reduced attitude is mainly induced by translational acceleration. However, fixed-wing UAVs are also used in practice and the dynamics are different. In future, one can extend our results to fixed-wing UAV systems, where aerodynamic constraints, bank-angle dynamics, and minimum-speed requirements need to be jointly considered. Secondly, the current study considered a single-antenna directional communication model. Future research may extend the proposed dynamics-consistent design to MIMO-enabled UAV covert communications, where trajectory, attitude, and transmit beamforming should be jointly optimized. Thirdly, although bounded attitude-control errors were considered, perfect CSI was assumed to be available. It is of interest to develop more robust designs under imperfect CSI, especially when the warden's channel or location information is uncertain.

\vspace{-2mm}
\bibliographystyle{IEEEtran}
\bibliography{IEEEcovert} 
\end{document}